\pdfoutput=1

\documentclass[reprint,amsmath,twocolumn,superscriptaddress,10pt,aps]{revtex4-2} % this article uses the Revtex 4-1 format

\usepackage{graphicx}
\usepackage{amsmath}
\usepackage{amssymb}
\usepackage{amsfonts}
\usepackage{bm}
\usepackage{xcolor}
\usepackage{hyperref}
\usepackage{setspace}
\usepackage[nearskip,margin=0pt,caption=false]{subfig}

\usepackage{verbatim}
\usepackage{epstopdf}
\usepackage{lipsum,babel}
\usepackage{siunitx}
\usepackage{array}
\usepackage[normalem]{ulem}
\usepackage{layouts}
\usepackage{etoolbox}
\usepackage{float}

\usepackage{array}
\usepackage{booktabs}

\usepackage[resetlabels,labeled]{multibib}
\newcites{SI}{Supplementary References}

\usepackage{xparse}

\let\origciteSI\citeSI

\RenewDocumentCommand{\citeSI}{o o m}{%
  {%
    \renewcommand{\citenumfont}[1]{SI##1}%
    \IfNoValueTF{#1}
      {\origciteSI{#3}}
      {%
        \IfNoValueTF{#2}
          {\origciteSI[#1]{#3}}
          {\origciteSI[#1][#2]{#3}}%
      }%
  }%
}

\DeclareGraphicsExtensions{.pdf,.eps,.png,.jpg,.mps}

\graphicspath{
  {Figures/}
  {figures/}
}
\begin{document}

\title{Integrated ytterbium gain for visible-near-infrared photonics}
% \vspace{-0.2cm}

\author{Tianyi~Zeng}
\thanks{These authors contributed equally to this work.}
\affiliation{John\ A.\ Paulson\ School\ of\ Engineering\ and\ Applied\ Sciences,\ Harvard\ University}

\author{Erik~W.~Masselink}
\thanks{These authors contributed equally to this work.}
\affiliation{John\ A.\ Paulson\ School\ of\ Engineering\ and\ Applied\ Sciences,\ Harvard\ University}
\affiliation{Department\ of\ Electrical\ Engineering\ and\ Computer\ Science,\ Massachusetts\ Institute\ of\ Technology}

\author{Tsung-Han~Wu}
\affiliation{Electrical,\ Computer\ and\ Energy\ Engineering,\ University\ of\ Colorado,\ Boulder}
\affiliation{Chi3\ Optics\ LLC,\ Boulder,\ CO,\ USA}

\author{Nathan~Brooks}
\affiliation{Chi3\ Optics\ LLC,\ Boulder,\ CO,\ USA}

\author{Peter~Chang}
\affiliation{Chi3\ Optics\ LLC,\ Boulder,\ CO,\ USA}

\author{Grisha~Spektor}
\affiliation{Octave\ Photonics,\ Louisville,\ CO,\ USA}

\author{Zachary~L.~Newman}
\affiliation{Octave\ Photonics,\ Louisville,\ CO,\ USA}

\author{Danxian~Liu}
\affiliation{John\ A.\ Paulson\ School\ of\ Engineering\ and\ Applied\ Sciences,\ Harvard\ University}

\author{Scott~B.~Papp}
\affiliation{Time\ and\ Frequency\ Division,\ National\ Institute\ of\ Standards\ and\ Technology,\ Boulder,\ CO,\ USA}
\affiliation{Department\ of\ Physics,\ University\ of\ Colorado\ Boulder}

\author{David~R.~Carlson}
\affiliation{Octave\ Photonics,\ Louisville,\ CO,\ USA}

\author{Scott~A.~Diddams}
\affiliation{Electrical,\ Computer\ and\ Energy\ Engineering,\ University\ of\ Colorado,\ Boulder}
\affiliation{Department\ of\ Physics,\ University\ of\ Colorado\ Boulder}

\author{Kiyoul~Yang}
\email{kiyoul@seas.harvard.edu}
\affiliation{John\ A.\ Paulson\ School\ of\ Engineering\ and\ Applied\ Sciences,\ Harvard\ University}

\begin{abstract}
\noindent Rare-earth gain media \cite{main:Desurvire:1987:OL,main:hanna:1988:EL:YDFAlaser} form the foundation of modern optical communications, emerging quantum hardware \cite{main:Doyle:2019:Science,main:Endres:2025:Nature}, and ultrafast optics \cite{main:Udem:2008:Science,main:papp:2018:science:subcycle,main:diddams:2021:NP:fewcycle:Er}. While chip-scale integration can enable fiber-like -- and potentially beyond-fiber -- functionality with unprecedented scalability, development in the visible and near-infrared remains in its early stages. Here, we demonstrate ytterbium-based optical gain integrated into an aluminum oxide photonic platform, achieving both single-mode lasing and optical amplification in the near-infrared regime. This platform delivers optical amplification with output powers exceeding 0.5 W, an optical-to-optical conversion efficiency above 70\%, and a noise figure of 3.3 dB, approaching the quantum limit for phase-insensitive amplification. Furthermore, we achieve femtosecond pulse amplification to a record peak power of 14 kW, enabling supercontinuum generation with visible dispersive waves extending from 780 to 476 nm in conjunction with nonlinear photonic devices. This platform is compatible with heterogeneous integration into standard photonic circuits, laying the foundation for scalable visible-near-infrared photonic systems, including coherent laser arrays, mode-locked lasers \cite{main:marandi:2023:science,main:Kippenberg:2025:arxiv}, optical clocks \cite{main:Schmidt:2015:Review_modern_physics,main:ACES:2019:Optica}, and microwave oscillators \cite{main:Gryphon:2024:Nature,main:quinlan:2026:NP:feedforward}.

\end{abstract}

\maketitle

% Do not write main-text section entries to the SI table of contents
\let\savedaddcontentsline\addcontentsline
\renewcommand{\addcontentsline}[3]{}

%%% PAPER STARTS HERE
\section{Introduction}

\begin{figure*}[htb!]
    \centering
    \includegraphics[width=0.95\linewidth]{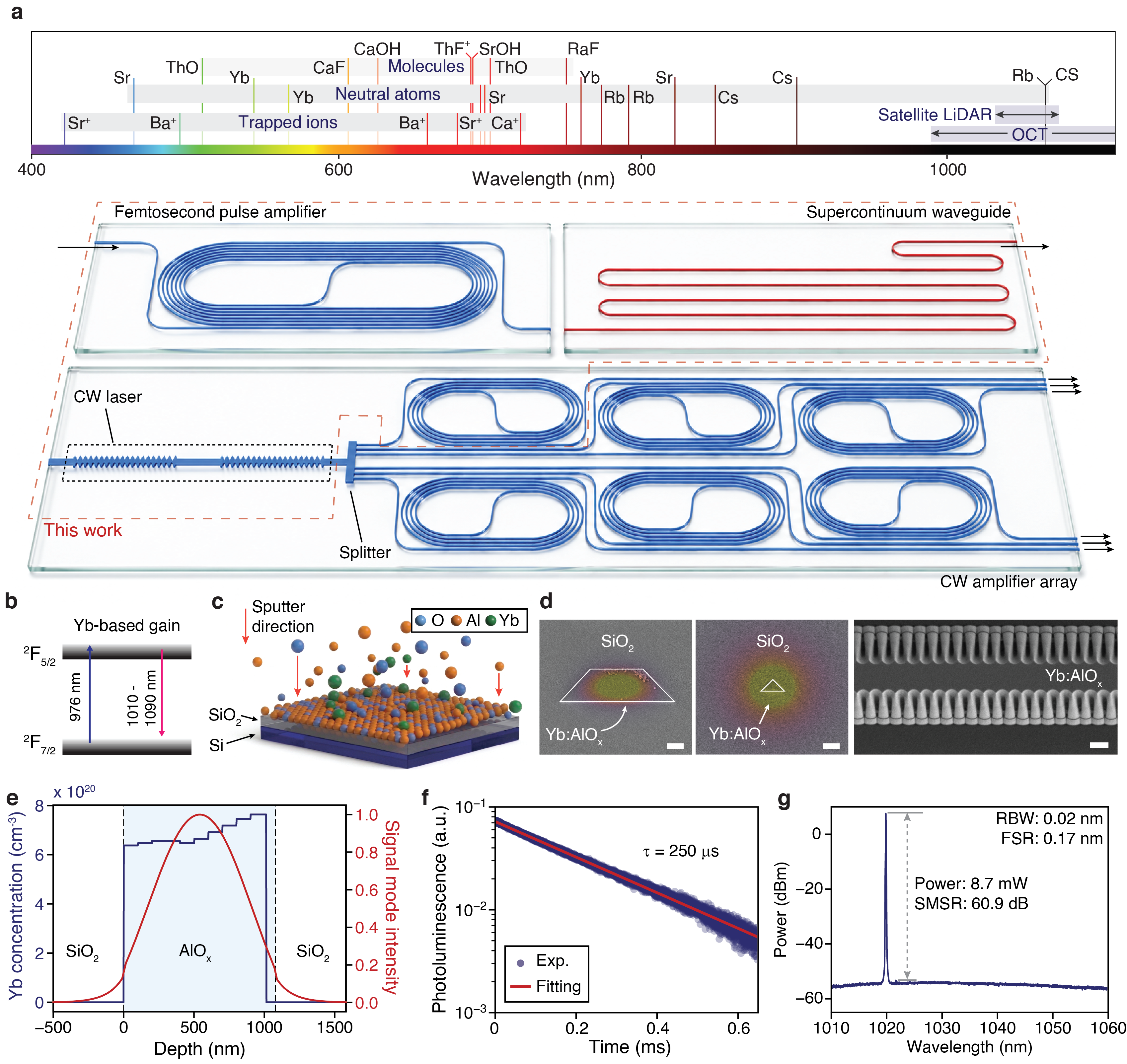}
    \caption{\textbf{Integrated ytterbium-doped alumina photonic platform.} (\textbf{a}) \textbf{Top:} Spectrum coverage in the visible to short NIR regions, illustrating wavelength requirements for various photonic applications and physical systems \cite{main:Doyle:2019:Science,main:Doyle:2024:Nature,main:Endres:2025:Nature,main:Lahaye:2020:Nature_Physics,main:Schmidt:2015:Review_modern_physics,main:Weiner:2025:Nature_Photonics}. %
    \textbf{Bottom:} Schematics of two photonic systems incorporating Yb:AlO$_x$ amplifiers. The first consists of a pulse amplifier followed by a nonlinear  waveguide for supercontinuum generation.
    The second system consists of a CW amplifier array seeded by a master single-mode laser.
    (\textbf{b}) A simplified energy-level diagram of the Yb$^{3+}$ ions in alumina, with pumping at 976 nm (as used in this work) and the signal wavelengths spanning from 1010 to 1090 nm.
    (\textbf{c}) Schematic of the co-sputtering process for depositing a Yb:AlO$_x$ film onto an oxidized silicon substrate. (\textbf{d}) SEM images, from left to right: cross-sectional images of the amplifier waveguide, amplifier facet and a top view of the Bragg reflector for the DFB laser. Simulations of the signal modes are overlaid. All scale bars in this panel are 500 nm. 
    (\textbf{e}) Vertical profiles of the measured dopant concentration (characterized by RBS) and the simulated optical TE$_{00}$-mode intensity across the waveguide center. (\textbf{f}) Measured photoluminescence decay from the $^{2}$F$_{5/2}$ manifold, with a fitted curve overlaid, showing a lifetime of 250~$\mu$s. (\textbf{g}) Measured output optical spectrum of a Yb:AlO$_x$ single-mode DFB laser. OSA resolution bandwidth (RBW): 0.02 nm; cavity free spectral range (FSR): 0.17 nm.
     }
    \label{fig:intro}
\end{figure*}

Rare-earth-based optical amplifiers and lasers \cite{main:Desurvire:1987:OL,main:hanna:1988:EL:YDFAlaser,main:Hanna:2002:IEEE_JQE} are foundational to modern science and technology, most notably serving as the backbone of global optical communications. They underpin a broad range of frontier scientific explorations (Fig. \ref{fig:intro}a) -- particularly at short near-infrared (NIR) and visible wavelengths -- including neutral-atom quantum processors \cite{main:Endres:2025:Nature,main:lukin:2025:Nature}, 
% ,Saffman:2022:Nature, lukin:2021:Nature:Rb809nmTweezer,
manipulation of ultracold molecules \cite{main:Doyle:2019:Science,main:Doyle:2024:Nature}, nonlinear microscopy in neuroscience \cite{main:Xu:2024:Cell}, 
% Xu:2013:NaturePhotonics, dunn:2016:BOE:YbML,
and precision spectroscopy in astronomical observatories \cite{main:Udem:2008:Science,main:Holzwarth:2012:Nature,main:diddams:2019:OSA:astrocomb}.
% notes Udem:2008 is 1583 nm, Erbium 
The rapid adoption of rare-earth-doped, particularly ytterbium-doped, fiber systems across these visible-NIR regimes is directly driven by their unparalleled performance metrics: high output power and high optical-to-optical 
conversion efficiency, low noise figure, broad gain bandwidth, and low thermal load. 
Incorporating rare-earth ions into thin-film waveguides has been widely explored
as scalable gain platforms, using a range of dopants including erbium, ytterbium, neodymium, and thulium, in host materials such as 
silica, aluminum oxide, lithium niobate, silicon nitride, tantalum pentoxide, and tellurium oxide 
\cite{main:polman:1993:APL:ErImplant,main:polman:1994:APL:ErLN:implant,main:vahala:2004:PRA:ErImplant:SiO2,main:Pollnau:2007:APB,main:watts:2014:OE:bradley:ErYb:ringlaser,main:blumenthal:2014:OE:ErAlO:DFBDBR,main:wilkinson:2015:OL:Ybtantala,main:bradley:2020:OE:codope:solubility:wetetch,main:bradley:2020:PR:ErTeO2SiN,main:kippenberg:2022:science,main:kartner:2025:NatPhoton,main:blanco:2026:OE:Nd,main:blanco:2025:Optics_express}.
However, as integrated photonics has expanded into the visible and NIR regimes \cite{main:Chiaverini:2020:Nature,main:Nexus:2022:Nature,main:Loncar:2023:NatureCommun,main:Vahala:2026:Nature}, the successful incorporation of rare-earth-based gain and light sources in this wavelength range remains a critical, unresolved challenge. 

This work presents the integration of Yb-based optical gain for visible-NIR photonics, leveraging two fundamental physical advantages. First, the simple energy-level structure of Yb$^{3+}$ ions (Fig. \ref{fig:intro}b) facilitates high output power and optical-to-optical conversion efficiency in the NIR \cite{main:Hanna:2002:IEEE_JQE}. The corresponding reduction in thermal load is critical, as heat dissipation and its associated thermorefractive noise remain primary bottlenecks in integrated photonics \cite{main:kippenberg:2019:PRA,main:hamerly:2020:PRX,main:papp:2020:OL:precise,main:gaeta:2026:NP}. Second, the Yb gain band (1010 nm - 1090 nm) lies near the edge of the anomalous dispersion window in integrated nonlinear platforms. Consequently, ultrafast pulses generated and amplified within this band can undergo spectral broadening to reach -- and potentially fully encompass -- the visible wavelength range. 
We integrate Yb$^{3+}$ ions into a low-loss aluminum oxide (AlO$_x$, hereafter alumina) platform \cite{main:polman:1993:APL:ErImplant,main:Pollnau:2007:APB,main:bradley:2024:JVSTA:lowTempAlO150C}, enabling the integration of a wide range of on-chip photonic components, 
material systems, as well as CMOS electronics \cite{main:campbell:2008:NP:Ge,main:stajanovic:2015:Nature,main:ram:2018:Nature,main:Nexus:2022:Nature,main:kippenberg:2023:NatComm:LNonSiN,main:papp:2026:Nature:TantalaLN}.

Using this platform, we demonstrate low-noise continuous-wave amplification with efficiencies exceeding 70\%, along with ultrafast pulse amplification to nanojoule-level energies and 14 kW peak power after compression. These high-energy pulses drive supercontinuum generation from the NIR into the blue, establishing an integrated route toward short NIR and visible light generation (Fig. \ref{fig:intro}a).  

\begin{figure*}[htb!]
    \centering
    \includegraphics[width=0.95\linewidth]{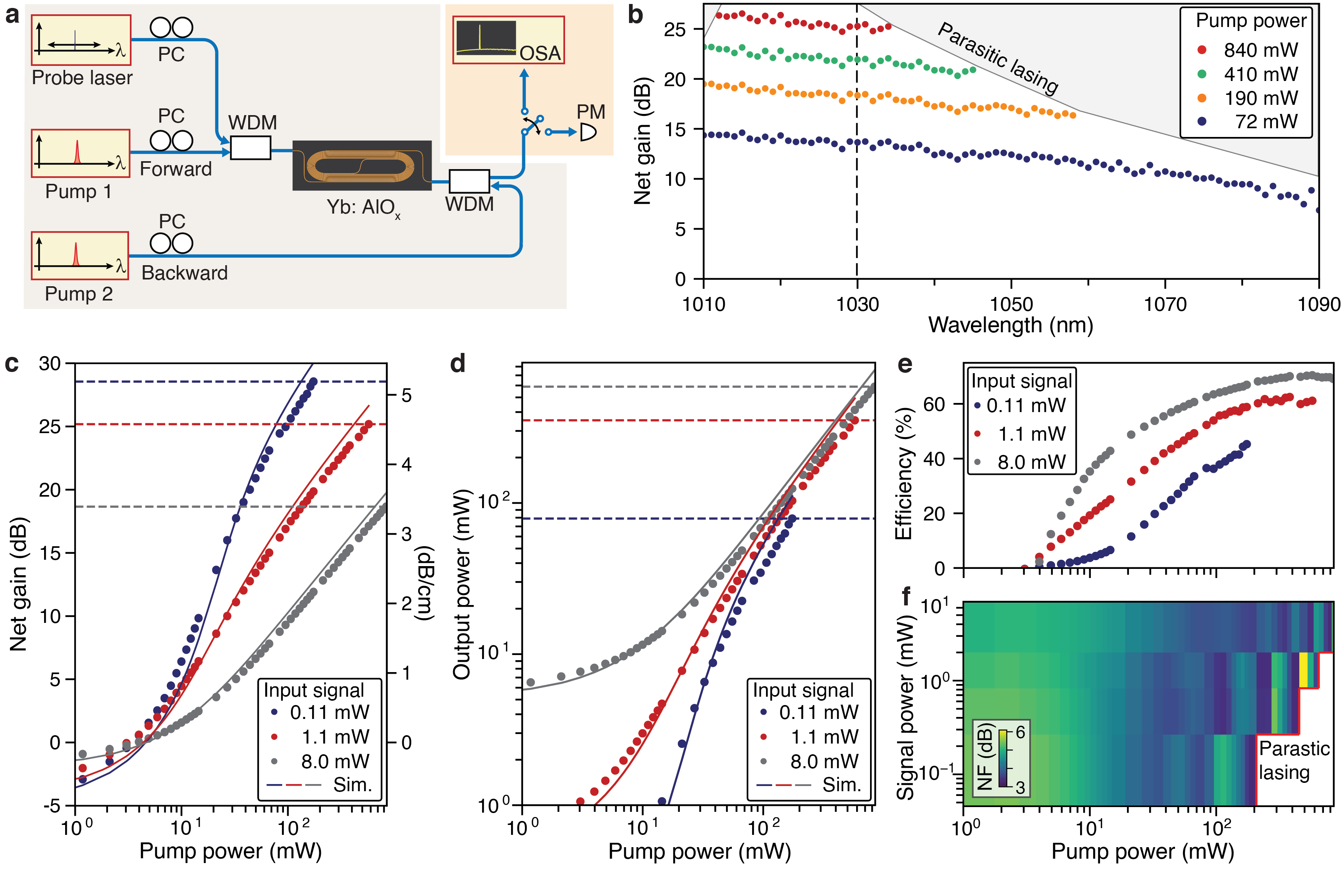}
    \caption{\textbf{CW optical amplification.} (\textbf{a}) Experimental setup for CW optical amplification with bidirectional pumping (pump 1 and 2) and a tunable probe laser. PC, polarization controller; OSA, optical spectrum analyzer; PM, power meter; WDM, wavelength division multiplexer. (\textbf{b}) Measured gain spectra at different pump powers; the gray shaded region marks the onset and parameter space of parasitic lasing. 
    (\textbf{c-d}) Measured data (points) and simulations (lines) of the on-chip net gain (\textbf{c}) and output power (\textbf{d}) at 1030 nm. (\textbf{e}) Measured optical-to-optical conversion efficiency as a function of pump power for different signal powers. 
    (\textbf{f}) Measured noise figure. The red contour indicates the onset of parasitic lasing. (\textbf{e-f}) share a common horizontal axis and were measured at 1030 nm. }
    \label{fig:CW}
\end{figure*}

\begin{figure*}[htb!]
    \centering
    \includegraphics[width=0.95\linewidth]{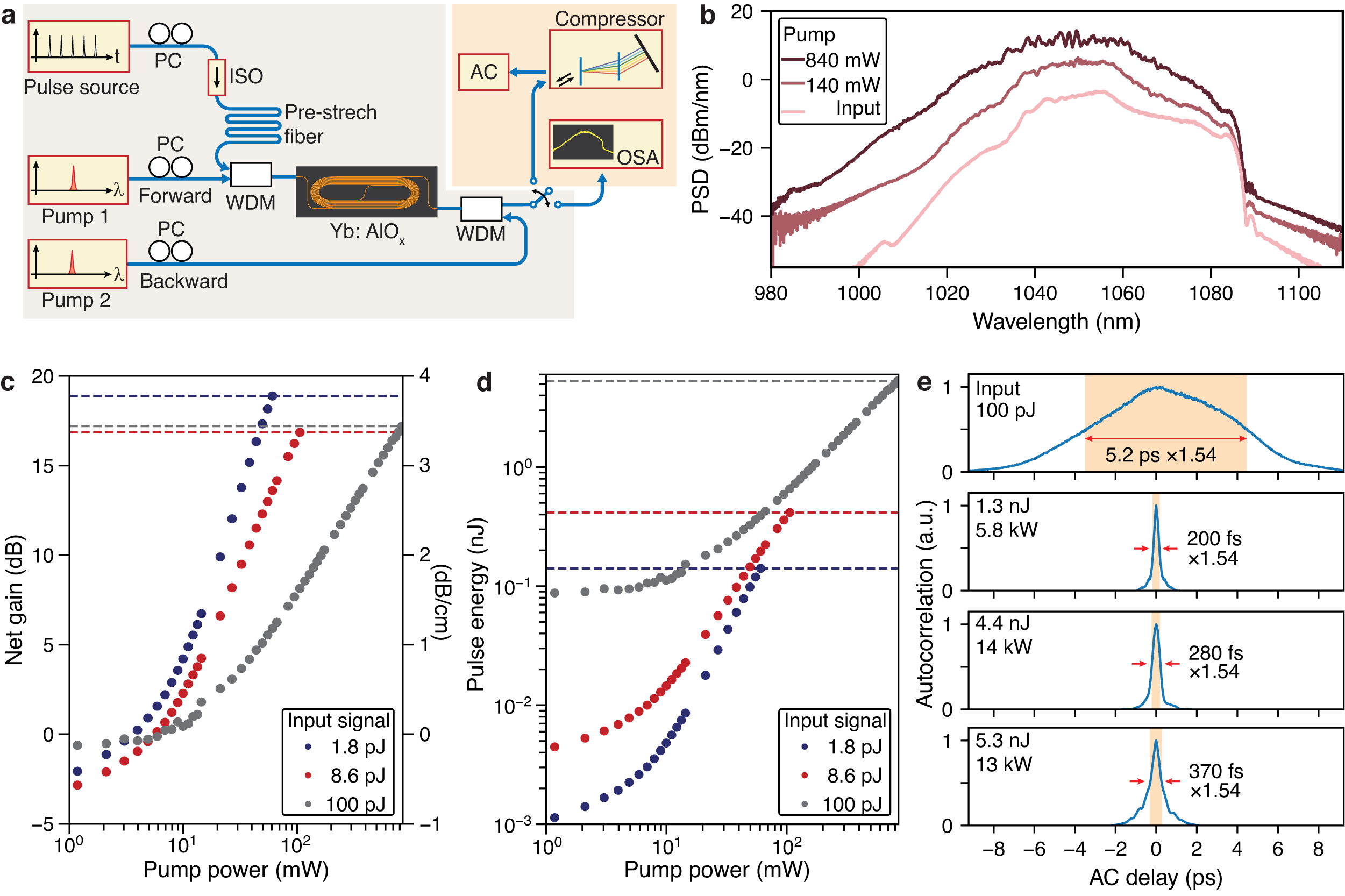}
    \caption{\textbf{Femtosecond pulse amplification.} (\textbf{a}) Experimental setup for femtosecond pulse amplification with bidirectional pumping and a fiber-based pulse source. AC, autocorrelator; ISO, isolator. 
    (\textbf{b}) Optical spectra of the on-chip input and amplified output at various pump powers for an input pulse energy of 100 pJ. (\textbf{c-d}) Measured net gain (\textbf{c}) and output pulse energy (\textbf{d}) at different input pulse energies. 
    (\textbf{e}) Optical intensity autocorrelation (AC) traces of the pre-chirped input pulse (top) and the amplified output pulses. 
    Pulse energies and peak powers are indicated in the upper-left corner of each subplot.
    The output traces were recorded after optimized linear chirp compression. The pulse widths were calculated using a deconvolution factor of 0.65, assuming a sech-squared pulse shape.
     All reported pulse energies and peak powers are on-chip. }
    \label{fig:pulse}
\end{figure*}

\begin{figure*}[htbp!]
    \centering
    \includegraphics[width=0.95\linewidth]{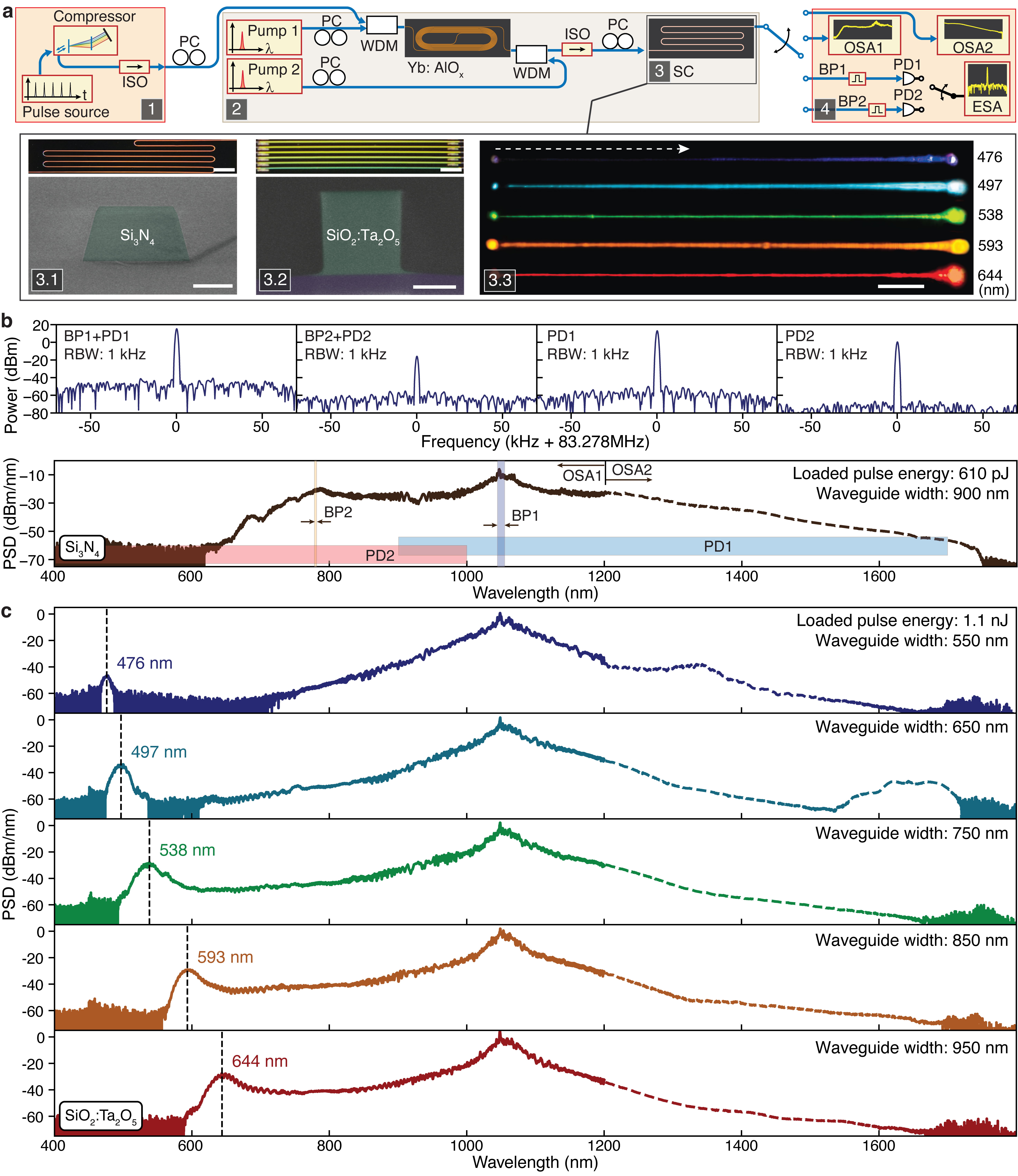}
    \caption{\textbf{Visible-to-NIR supercontinuum generation.} (\textbf{a}) Experimental setup for supercontinuum generation: (1) Fiber-based pulse source with a free-space compressor and isolator for pre-chirped input. (2) Bidirectionally pumped femtosecond pulse amplifier. (3) Supercontinuum (SC) waveguide. (4) Output characterization setup including OSAs, photodetectors (PD) with bandpass filters (BP), and an electrical spectrum analyzer (ESA). 
    Insets of (3): Optical top view and SEM cross-sections of Si$_3$N$_4$ (3.1) and $\mathrm{SiO_2\!:\!Ta_2O_5}$ (3.2) waveguides, and a top-view photograph of waveguide arrays showing visible dispersive-wave generation in the $\mathrm{SiO_2\!:\!Ta_2O_5}$ waveguides (3.3).
    Scale bars: 500 nm for the SEM images in panels (3.1) and (3.2), and 500 $\mu$m for all top views in panels (3.1)--(3.3).
    (\textbf{b}) Si$_3$N$_4$ supercontinuum spectrum, with narrow-band near-infrared (near-visible) comb lines filtered by BP1 (BP2) and detected by PD1 (PD2) for the repetition beatnotes shown above. Repetition beatnotes from the broadband optical signal measured by PD1 and PD2 without filters are also shown. 
    (\textbf{c}) Supercontinuum spectra of the $\mathrm{SiO_2\!:\!Ta_2O_5}$ waveguides with varying waveguide widths, showing the corresponding dispersive wave positions.
    All optical spectra were measured using two OSAs covering the full wavelength range of interest and stitched at 1200 nm.
    }
    \label{fig:supercontinuum}
\end{figure*}

\section{Design and Fabrication}

Alumina is well-suited as a host material due to its high rare-earth solubility \cite{main:bradley:2020:OE:codope:solubility:wetetch}, wide bandgap, and low optical nonlinearity \cite{main:mansoor:1996:IEEEJQE:alOn2,main:Ballota:2018:App_Glass}. To the best of our knowledge, amorphous alumina films have not been reported to exhibit photorefractive effects, in contrast to crystalline materials \cite{main:Kip:1998:ApplPhyB,main:Wilkinson:1995:OL}, consistent with our observation in this work. 
We employ the co-sputtering technique (Fig. \ref{fig:intro}c) that has been established for rare-earth-doped alumina photonic circuits \cite{main:Pollnau:2007:APB,main:bradley:2020:OE:codope:solubility:wetetch,main:yang:2025:CLEO:ErInjection,main:blanco:2026:OE:Nd}.
The highest process temperature in this work is $725~^\circ$C, used to anneal the top cladding rather than the alumina film. However, sputtered alumina has been demonstrated with processing temperatures as low as $150~^\circ$C \cite{main:bradley:2024:JVSTA:lowTempAlO150C}, potentially enabling compatibility with back-end-of-line passive photonics foundry processes \cite{main:blanco:2019:IEEE:AlOSiNEr}.

The co-sputtering process enables uniform dopant incorporation throughout the 1.08-$\mu$m-thick waveguide cores without inducing surface damage, resulting in a root-mean-square top-surface roughness of 120~pm, as measured by atomic force microscopy. After film growth, the waveguides were patterned lithographically and etched with reactive ion etching, followed by top-cladding deposition and annealing. The passive waveguide loss is 9.7~dB/m at 1060~nm and is dominated by etch-induced sidewall roughness. To reduce reflections and fiber-coupling loss, the waveguides were designed with 3D inverse tapers in both height and width, implemented in a single lithography and etching step, yielding a coupling loss of 1.0~dB per facet (characterization of passive and coupling loss in Supplementary Information). Cross-sectional scanning electron microscope (SEM) images are shown in Fig. \ref{fig:intro}d.

The resulting film was characterized by Rutherford backscattering spectrometry (RBS; Fig.~\ref{fig:intro}e) and exhibits a $\mathrm{Yb^{3+}}$ concentration of $6.8\times10^{20}\ \mathrm{atoms/cm^3}$, corresponding to 0.76~at.\% (atomic percentage) and 62{,}000~ppm by weight in the doped alumina layer.
The high concentration is enabled by the low quenching of $\mathrm{Yb^{3+}}$ due to its simple energy-level structure \cite{main:Hanna:2002:IEEE_JQE}. The luminescence lifetime of the $\mathrm{Yb^{3+}}$ ions was measured to be 250 $\mu$s by intensity-modulating the pump and monitoring the amplified spontaneous emission from the ytterbium-doped alumina waveguide (Fig. \ref{fig:intro}f). The lifetime is shorter than the typical $\mathrm{~1}$~ms lifetime observed in Yb-doped fibers \cite{main:Hanna:1997:OC:quenchingYDFA} due to the high doping concentration, but agrees with recent measurements in highly doped sputtered alumina \cite{main:pollnau:2025:OpticalMaterials:YbAlO:lifetime:quenching}.

The Yb-doped alumina waveguides are clad with silica (SiO$_2$) and are designed with a height of 1.08~\(\mu\)m, a width of 1.50~\(\mu\)m, and a \ang{52} sidewall angle, resulting in a mode area of 1.90~\(\mu\mathrm{m}^2\). This geometry leads to a high pump-signal mode overlap of 97.7\% and a doped region overlap factor of 89.5\%. The doped region overlap factor denotes the fraction of the signal mode contained in the doped region \cite{main:Watts:2013:OL}, and is calculated using the dopant profile measured by RBS (Fig. \ref{fig:intro}e). Together, the large mode area, high doped region overlap factor, and highly confined mode enable low nonlinearity, high gain, and a compact, scalable footprint, respectively.

As a proof of concept, single-mode operation of a continuous-wave (CW) laser is demonstrated in Fig. \ref{fig:intro}g. The distributed feedback (DFB) laser consists of two Bragg reflectors (Fig. \ref{fig:intro}d) providing a 0.7-nm-wide reflection band centered at 1020~nm, with a Yb-doped waveguide between them.  The laser cavity's free spectral range is 0.17 nm, extracted from the resonance spacing measured using a swept tunable laser. The device emits from both facets and delivers 8.7 mW of on-chip power per side, with a side-mode suppression ratio (SMSR) of 60.9 dB (see Supplementary Information).
\section{Continuous-wave amplification}

A 55-mm spiral amplifier was characterized under bidirectional 976-nm pumping (Fig. \ref{fig:CW}a). The pump and signal were combined for amplification and subsequently separated using 980/1060~nm wavelength-division multiplexers (WDMs), and chip coupling was performed with ultra-high-numerical-aperture fibers (Nufern UHNA3). We measured the spectral response of the amplifier by sweeping the wavelength of the signal laser, keeping its on-chip power fixed at 1.7 mW. Optical gain was observed from 1010 to 1090 nm, with a peak net gain of 26.5 dB at 1015 nm (Fig. \ref{fig:CW}b). Measurements of
higher gain were limited by parasitic lasing arising from
facet reflections.

We further characterized CW amplification at a fixed wavelength of 1030~nm (Fig. \ref{fig:CW}c-f). This wavelength was selected to minimize parasitic lasing and reduce insertion loss in the fiber WDM. It is therefore not chosen for maximum gain. For a small-signal input power of 110~$\mu$W at 1030~nm, we measured an on-chip net gain of 28.6~dB, corresponding to an off-chip net gain of 26.6~dB, accounting for the fiber-to-chip coupling loss of 1.0~dB per facet (Fig. \ref{fig:CW}c). The resulting gain coefficient is 5.2~dB/cm. In the large-signal regime, with an input power of 8.0~mW and a pump power of 840~mW, the maximum output power reaches 590~mW (Fig. \ref{fig:CW}d), corresponding to a gain of 18.7~dB. A maximum efficiency of 71\% is achieved at a lower pump power of 580~mW (Fig. \ref{fig:CW}e). The accompanying simulations are based on a two-level rate-equation model and agree qualitatively with our measurements (Fig. \ref{fig:CW}c--d). We used the model to extract an energy-transfer upconversion factor of $\mathrm{5\times 10^{-18}~cm^3/s}$ and to calculate the minimum spontaneous emission factor and saturation energy discussed below.

We also measured the noise figure (NF), observing a minimum off-chip value of 4.3~dB that corresponds to an on-chip value of 3.3~dB (Fig. \ref{fig:CW}f). This value is very close to the 3.29~dB limit calculated from the minimum spontaneous emission factor ({$n_{\mathrm{sp}}$}) of $\mathrm{Yb^{3+}}$. We note that asymmetric chip-to-fiber coupling at the two facets can bias the on-chip noise figure estimation.
A discussion of coupling asymmetry, the rate equation model, and the minimum noise figure is provided in Supplementary Information. 

\section{Ultrafast pulse amplification}

The Yb-doped alumina platform is well suited for pulse amplification, leveraging low material nonlinearity, large saturation energy, and long upper-state lifetime. 
The nonlinear index (n$_2$) of alumina is $4.8\times10^{-20}\ \mathrm{m^2/W}$ \cite{main:Ballota:2018:App_Glass}, which is comparable to that of silica and approximately 5--10 times lower than that of silicon nitride \cite{main:Vahala:2022:NatureCommun}. Combined with the short amplifier length of 5.5~cm, this minimizes the nonlinear phase accumulation. Additionally, we calculated a large saturation energy of 730~nJ (see Supplementary Information), such that each pulse depletes only a small fraction of the population inversion and therefore experiences nearly constant gain throughout the pulse, ensuring negligible distortion and gain depletion. 

These favorable properties enable low-distortion, energetic ultrafast pulse amplification of a Yb-fiber mode-locked laser emitting at a center wavelength of 1050~nm. 
The initial 125~fs pulses were first pre-chirped to 5.2~ps in normally dispersive HI 1060 fiber to reduce the peak power and mitigate nonlinear optical effects \cite{main:Mourou:1985:OptCommun}, then amplified in the integrated Yb amplifier, and finally re-compressed using a grating compressor (Fig. \ref{fig:pulse}a). The input and  output optical spectra are shown in Fig. \ref{fig:pulse}b and the average power was measured using a calibrated optical spectrum analyzer (OSA). For low-energy input pulses of 1.8~pJ, we measured a net gain of 18.9~dB (Fig. \ref{fig:pulse}c). 
This value is lower than the small-signal gain for CW amplification, because the pulse spectrum is not centered within the amplifier’s peak gain band of 1010–1030~nm (Fig. \ref{fig:CW}b). Increasing the input pulse energy to 100 pJ yields record-high output pulse energies of 5.3~nJ (Fig. \ref{fig:pulse}d). At output energies above 1~nJ, however, the compressed pulse duration begins to broaden, increasing from 200~fs to 370~fs (Fig. \ref{fig:pulse}e), due to nonlinear chirp accumulated in the amplifier that cannot be compensated by the grating compressor \cite{main:Agrawal:2000:Springer}.
A maximum peak power of 14~kW after compression is achieved at an output pulse energy of 4.4~nJ, due to the narrower pulse width. Given the long upper-state lifetime of 250~$\mu$s, reducing the pulse repetition rate could further increase the output pulse energy. 
In practice, however, the achievable pulse energy may be limited by nonlinear distortions, as observed here. These distortions can be further mitigated by increasing the mode area or by balancing dispersion, gain, and nonlinearity to approach a self-similar pulse-evolution regime \cite{main:Harvey:2000:Phys_Rev_Let}.

\section{Supercontinuum generation}

After amplification in the Yb-doped alumina amplifier, femtosecond pulses with energies of 2.6 nJ (measured in fiber) were injected into silicon nitride ($\mathrm{Si_3N_4}$) and silica-tantala ($\mathrm{SiO_2\!:\!Ta_2O_5}$) supercontinuum waveguides \cite{main:papp:2020:OL,main:papp:2021:Optica,main:papp:2026:Nature:TantalaLN} (Fig. \ref{fig:supercontinuum}a(3.1) and \ref{fig:supercontinuum}a(3.2)). The silicon nitride waveguide, measuring 700~nm in thickness, 900~nm in width, and 50~mm in length, was designed to exhibit anomalous dispersion at 1050~nm and to generate a visible dispersive wave near 765~nm (integrated dispersion in Supplementary Information).
This dispersion enables a supercontinuum spectrum spanning approximately 600 to 1700~nm at a loaded pulse energy of 610~pJ (Fig. \ref{fig:supercontinuum}b). 

Repetition-rate beatnotes measured across the spectrum confirm that the supercontinuum remains coherent over the full bandwidth (Fig. \ref{fig:supercontinuum}b). 
For silica-tantala waveguides, the material dispersion was optimized by varying the material composition \cite{main:papp:2025:arxiv}, while the geometric dispersion was tuned by sweeping waveguide widths from 550 to 950 nm. 
At loaded pulse energies of 1.1 nJ, these waveguides generate supercontinua with dispersive wave locations spanning from red to the blue (Fig. \ref{fig:supercontinuum}a(3.3) and Fig. \ref{fig:supercontinuum}c). 
 
Notably, the green line at 538~nm is well-suited for f-2f locking, enabling stabilized frequency combs \cite{main:Vahala:2017:Nature_com}.

In this experiment, we used discrete Yb-doped amplifier and supercontinuum waveguide chips together with fiber-based WDMs and isolators. In a fully integrated system, the WDMs can be implemented on chip, and the isolator can be eliminated when the Yb-doped amplifier is heterogeneously integrated with the nonlinear photonic circuit.  Bidirectional pumping can be implemented via additional pump-injection waveguides with on-chip WDMs or directly through the nonlinear photonic waveguide.
\section{Discussion and outlook}

We have demonstrated an integrated Yb-doped gain platform with a small-signal gain of 28.6~dB, greater than 0.5~W output power, and an optical-to-optical efficiency over 70\%. The broad gain bandwidth (1010--1090~nm), together with the low nonlinearity, long lifetime, and large saturation energy, enables pulsed amplification to record-high pulse energies of 5.3~nJ and compressed peak powers of 14~kW. The high pulse energies drive integrated supercontinuum generation spanning from the NIR deep into the blue wavelengths.

Direct f-2f locking becomes feasible when combined with heterogeneously integrated thin film lithium niobate \cite{main:papp:2026:Nature:TantalaLN} or effective $\chi^{(2)}$ processes in silicon nitride \cite{main:Srinivasan:2021:NaturePhotonics,main:Bres:2022:NaturePhotonics,main:Vahala:2025:Science_advances}. Looking forward, full integration of a pulse generator can be realized on the Yb-gain platform established in this work via a Mamyshev oscillator \cite{main:Wise:2017:Optica,main:Kippenberg:2025:arxiv}, a passive Kerr soliton cavity \cite{main:kippenberg:2014:NP}, an electro-optic modulator \cite{main:marandi:2023:science}, or a saturable absorber for mode locking \cite{main:Keller:1996:IEEE_JSTQE,main:kartner:2009:IEEE:fs}. Furthermore, dispersion engineering enables pulse pre-chirping -- in both anomalous and normal regimes -- within nanophotonic waveguides, offering a direct on-chip integration of the free-space compressor utilized in this study. This line of development establishes a foundational architecture for optical frequency synthesizers \cite{main:DODOS:2018:Nature}, optical clocks \cite{main:ACES:2019:Optica}, and microwave oscillators \cite{main:Gryphon:2024:Nature,main:quinlan:2026:NP:feedforward} operating directly in the visible-NIR regime.  

Beyond the visible-NIR ultrafast photonics, this platform also provides a highly scalable architecture for producing multi-output, multi-watt CW light near 1~$\mu$m via an amplifier array \cite{main:kippenberg:2024:OFC:multilane} seeded by a single laser. When integrated with beam emitter arrays, this capability promises to enable scalable optical tweezer arrays for Cs \cite{main:Endres:2025:Nature}, % ,Saffman:2022:Nature
along with other quantum computing applications. Furthermore, it establishes a foundation for nonlinear frequency conversion capable of providing the critical visible and ultraviolet cooling lasers required for ultracold atoms, ions, and molecules \cite{main:Schmidt:2015:Review_modern_physics,main:Doyle:2019:Science,main:Doyle:2024:Nature}. Finally, the co-sputtering approach demonstrated here is not restricted to specific host materials or dopants. For instance,
co-doping Er with Yb can significantly increase the pump absorption efficiency \cite{main:Polman:2003:Optical_materials,main:bradley:2020:OE:codope:solubility:wetetch}, enhancing the suitability for high-power amplification in the telecommunication band, pushing performance beyond currently existing limits.
\vspace{-0.1in}
\section*{Acknowledgments}
\vspace{-0.1in}
Major funding for this work is provided by the DARPA (D23AP00251-00), along with supports from the Office of the Under Secretary of Defense for Research and Engineering (FA8721-05-C-0002), and Harvard University. S.A.D. acknowledges support from AFOSR (FA9550-22-1-0483) and NSF (OMA-2016244). T.Z. acknowledges postdoctoral fellowship support from the Gordon and Betty Moore Foundation.

The alumina and silicon nitride devices were fabricated in the Harvard Center for Nanoscale Systems and MIT.nano cleanrooms. We thank J. Deng and E. Macomber for helpful discussions. We acknowledge Martin Chicoine (Laboratoire René-J-A-Lévesque, Université de Montréal) for conducting the Rutherford backscattering spectrometry measurements. We thank T. P. Letsou, F. Capasso, X. Fan, G. Semeghini, J. Doyle (Harvard), P. Delfyett (CREOL), W. Loh, D. Gray (MIT Lincoln Laboratory), M. Tran, T. Komljenovic (Nexus Photonics), Y. Wan (KAUST), and K. Van Gasse (Ghent) for discussions and/or internal review. K.Y. acknowledges discussions with J. Cohen (DARPA) and M. Lipson (Columbia). 

Any subjective opinions, interpretations, conclusions that might be expressed in the paper do not necessarily represent the views of the US Government.\\

% \noindent \textbf{Author contributions} T.Z., E.W.M. and K.Y. conceived the project. T.Z. and E.W.M. designed and fabricated the integrated amplifiers, and D.L. assisted with fabrication. T.W., N.B., P.C. and S.A.D. constructed the mode-locked laser and advised on compressor design. E.W.M. designed and T.Z. fabricated the silicon nitride supercontinuum waveguides. G.S., Z.N., S.B.P. and D.R.C. designed and fabricated the tantala supercontinuum waveguides. E.W.M. and T.Z. performed the experiments. T.Z., E.W.M., and K.Y. analyzed the data and wrote the paper with inputs from all authors. K.Y. supervised the project.\\

\noindent\textbf{Data availability}  The data that support the plots within this paper and other findings of this study are available from the corresponding author upon reasonable request.\\

\noindent\textbf{Code availability}  The code used to produce the plots within this paper is available from the corresponding author upon reasonable request.\\

\noindent\textbf{Conflict of interest} P.C., N.B. and T.-H.W. are affiliated with Chi3 Optics, which develops fiber lasers and integrated frequency comb systems. D.R.C. and Z.L.N. are affiliated with, and hold an ownership interest in, Octave Photonics, which manufactures and provides nanophotonic devices and services.

\bibliographystyle{naturemag}
\bibliography{Reference_Main}
% Re-enable TOC entries for SI
\let\addcontentsline\savedaddcontentsline
\newpage

\clearpage
\onecolumngrid

% \begin{center}
% {\Large \textbf{Supplementary Information}}
% \end{center}

\vspace{1em}

% Reset SI counters
\setcounter{section}{0}
\setcounter{subsection}{0}
\setcounter{figure}{0}
\setcounter{table}{0}
\setcounter{equation}{0}

% SI numbering style
\renewcommand{\thesection}{S\arabic{section}}
\renewcommand{\thesubsection}{S\arabic{section}.\arabic{subsection}}
\renewcommand{\thefigure}{S\arabic{figure}}
\renewcommand{\thetable}{S\arabic{table}}
\renewcommand{\theequation}{S\arabic{equation}}

\begin{center}
{\Large \textbf{Supplementary Information for:}}\\[0.5em]
{\Large \textbf{Integrated ytterbium gain for visible-near-infrared photonics}}\\[1.5em]

Tianyi Zeng$^{1,\dagger}$, Erik W. Masselink$^{1,2,\dagger}$, Tsung-Han Wu$^{3,4}$, Nathan Brooks$^{4}$, Peter Chang$^{4}$, Grisha Spektor$^{5}$,\\
Zachary L. Newman$^{5}$, Danxian~Liu$^{1}$, Scott B. Papp$^{6,7}$, David R. Carlson$^{5}$, Scott A. Diddams$^{3,7}$, Kiyoul Yang$^{1,*}$\\[0.8em]

{\small
$^1$John A. Paulson School of Engineering and Applied Sciences, Harvard University\\
$^2$Department of Electrical Engineering and Computer Science, Massachusetts Institute of Technology\\
$^3$Electrical, Computer and Energy Engineering, University of Colorado Boulder\\
$^4$Chi3 Optics LLC, Boulder, CO, USA\\
$^5$Octave Photonics, Louisville, CO, USA\\
$^6$Time and Frequency Division, National Institute of Standards and Technology, Boulder, CO, USA\\
$^7$Department of Physics, University of Colorado Boulder\\
$^\dagger$These authors contributed equally to this work.\\
$^*$kiyoul@seas.harvard.edu
}
\end{center}

\vspace{1em}
\renewcommand{\thefigure}{S\arabic{figure}}
\setcounter{figure}{0}
\setcitestyle{numbers,square}

% Arabic section numbers
\makeatletter
\renewcommand\thesection{S\arabic{section}}
\renewcommand\thesubsection{S\arabic{section}.\arabic{subsection}}
\renewcommand\thesubsubsection{S\arabic{section}.\arabic{subsection}.\arabic{subsubsection}}

\renewcommand\p@subsection{}
\renewcommand\p@subsubsection{}
% --------------------------------------

\def\@hangfrom@section#1#2#3{\@hangfrom{#1#2}#3}
\def\@hangfroms@section#1#2{#1#2}
\makeatother

\onecolumngrid 
\setcounter{tocdepth}{2}
\tableofcontents
\newpage

\section{Methods}
\subsection{Numerical modeling}
\label{sec_modeling}
\begin{figure}[H]
  \centering
  \includegraphics[width=1\linewidth]{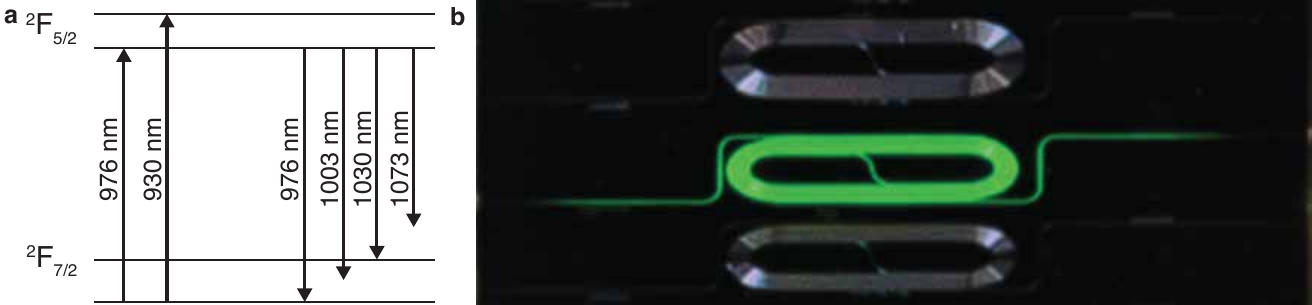}
  \caption{\textbf{Yb gain modeling. (a) }$\mathrm{Yb^{3+}}$ energy diagram in an alumina host \citeSI{Pollnau:2025:Opt_mat}. \textbf{(b)} Picture of the Yb amplifier showing green luminescence quenching.}
  \label{fig:YbLevels}
\end{figure}

The ytterbium-doped gain medium is modeled as an effective two-level system consisting of the ${}^2F_{7/2}$ ground-state manifold and the ${}^2F_{5/2}$ excited-state manifold (Fig. \ref{fig:YbLevels}a). The ground-state manifold is denoted as level~1 and the metastable upper laser manifold as level~2. We pump at 976~nm, corresponding to the lowest Stark levels of the ${}^2F_{5/2}$ manifold, because this wavelength provides a large absorption cross section and a relatively small quantum defect compared with shorter-wavelength Yb pump bands. 
Although Yb has been demonstrated to be predominantly inhomogeneously broadened in Al-doped silica \citeSI{Henry:2011:APB} and is therefore likely also inhomogeneously broadened in the alumina host, we treat the ions as a single effective species described by ensemble-averaged cross sections, which is appropriate  when operating at a single wavelength.

The population dynamics are governed by
\begin{equation}
\begin{gathered}
\begin{aligned}
\frac{dN_2}{dt} &= R_{12}N_1 - R_{21}N_2 - A_{21}N_2 - 2C_{\mathrm{ETU}}N_2^2, \\
\frac{dN_1}{dt} &= -R_{12}N_1 + R_{21}N_2 + A_{21}N_2 + 2C_{\mathrm{ETU}}N_2^2,
\end{aligned}
\\[4pt]
N_1 + N_2 = N_{\mathrm{Yb}},
\end{gathered}
\label{eq:rate}
\end{equation}
where $N_i$ is the population density of level $i$ and $N_{\mathrm{Yb}}$ is the average Yb concentration in the film measured with RBS (see Sec. \ref{sec_RBS}). The spontaneous decay rate is given by $A_{21}=1/\tau$ (see Sec. \ref{sec_Lifetime}), and the stimulated transition rates are the sum of the stimulated emission rates from each frequency, written as

\begin{equation}
\begin{gathered}
R_{ij} = \sum_k R_{ij,k}, \\
R_{ij,k} = \sigma_{ij,k}\phi_k = \frac{\sigma_{ij,k}P_k}{A_{\mathrm{eff},k} h \nu_k},
\end{gathered}
\end{equation}
where $\sigma_{ij}$ is the absorption or emission cross section, $\phi$ is the photon flux density, $h$ is the Planck constant, and $\nu_k$ is the frequency \citeSI{kippenberg:2022:Science}. The effective mode area $A_{\mathrm{eff},k}$ at the pump and signal frequencies was determined to be 1.85 $\mu \mathrm{m^2}$ and 1.90 $\mu \mathrm{m^2}$, respectively.

In contrast to rare-earth ions such as erbium or thulium, isolated Yb$^{3+}$ does not possess  higher-lying excited manifolds that strongly contribute to parasitic quenching. At high doping concentrations, however, Yb ion pairs and possible Er contamination can introduce additional loss channels, including energy-transfer upconversion (ETU) and potentially excited-state absorption (ESA), by enabling access to higher-energy states \citeSI{Pollnau:2025:Opt_mat}. The resulting green emission is seen in Fig. \ref{fig:YbLevels}b. In the present model, these effects are represented phenomenologically by a single homogeneous ETU coefficient, $C_{\mathrm{ETU}}$, which was fit to the amplifier data to be $\mathrm{5\times 10^{-18}~cm^3/s}$. Alternatively, separating the Yb population into paired and unpaired species with different ETU coefficients or including excited-state absorption, produces similarly good agreement.

The evolution of the pump, signal, and amplified spontaneous emission (ASE) powers along the waveguide is described by
\begin{equation}
\begin{aligned}
\frac{dP_s(z)}{dz} &= \Gamma_s\left[\sigma_{21,s}N_2(z)-\sigma_{12,s}N_1(z)\right]P_s(z)-\alpha_{0,s}P_s(z), \\
\frac{dP_{p,f}(z)}{dz} &= \Gamma_p\left[\sigma_{21,p}N_2(z)-\sigma_{12,p}N_1(z)\right]P_{p,f}(z)-\alpha_{0,p}P_{p,f}(z), \\
\frac{dP_{p,b}(z)}{dz} &= -\Gamma_p\left[\sigma_{21,p}N_2(z)-\sigma_{12,p}N_1(z)\right]P_{p,b}(z)+\alpha_{0,p}P_{p,b}(z), \\
\frac{dP_{\mathrm{ASE},f}(z)}{dz} &= \Gamma_{\mathrm{ASE}}\left[\sigma_{21,\mathrm{ASE}}N_2(z)-\sigma_{12,\mathrm{ASE}}N_1(z)\right]P_{\mathrm{ASE},f}(z) \\
&\quad + \Gamma_{ASE}P_{\mathrm{ASE}}^{0}\sigma_{21,\mathrm{ASE}}N_2(z)-\alpha_{0,\mathrm{ASE}}P_{\mathrm{ASE},f}(z), \\
\frac{dP_{\mathrm{ASE},b}(z)}{dz} &= -\Gamma_{\mathrm{ASE}}\left[\sigma_{21,\mathrm{ASE}}N_2(z)-\sigma_{12,\mathrm{ASE}}N_1(z)\right]P_{\mathrm{ASE},b}(z) \\
&\quad - \Gamma_{\mathrm{ASE}}P_{\mathrm{ASE}}^{0}\sigma_{21,\mathrm{ASE}}N_2(z)+\alpha_{0,\mathrm{ASE}}P_{\mathrm{ASE},b}(z),
\end{aligned}
\label{eq:propagation}
\end{equation}
where $P_s$, $P_p$, and $P_{\mathrm{ASE}}$ denote the signal, pump, and ASE powers, respectively, and the subscripts $f$ and $b$ denote forward- and backward-propagating waves. The doped region overlap factors $\Gamma_s = \Gamma_{ASE} = 89.5\%$ and $\Gamma_p = 90.2\%$ account for the spatial overlap of the optical mode in the doped region and are defined as $\Gamma_{s/p} = (\int_{active}I_{s/p}dA)/(\int_{\infty}I_{s/p}dA)$ \citeSI{Watts:2013:Optica}. The signal and pump modes are strongly co-located, with a calculated mutual overlap of 97.7\%. The mode intensities were calculated using a finite element mode solver (Tidy3D). Approximating the full transverse distributions by power-independent overlap factors is less accurate near threshold, but becomes accurate in the high-pump, unsaturated-gain regime \citeSI{Desurvire:1995:Book}. The passive propagation  loss is denoted as $\alpha_0$ (see Sec. \ref{sec_loss}).

Spontaneous emission contributes locally to the ASE power through the source term
\begin{equation}
P_{\mathrm{ASE}}^{0}=2h\nu_{\mathrm{ASE}}B,
\end{equation}
where $h$ is Planck’s constant, $\nu_{\mathrm{ASE}}$ is the ASE center frequency, and $B$ is the effective optical bandwidth \citeSI{Desurvire:1995:Book}. In this model, ASE is treated as a single-frequency channel, where the bandwidth $B\approx 14~\mathrm{THz}$ is used to represent the main Yb gain bandwidth from 1010--1060~nm. The coupled propagation equations were solved self-consistently with boundary conditions given by the input signal and bidirectional pump powers, with \(P_{s}(0)=P_{s,\mathrm{in}}\), \(P_{p,f}(0)=P_{p,f,\mathrm{in}}\), \(P_{p,b}(L)=P_{p,b,\mathrm{in}}\), \(P_{\mathrm{ASE},f}(0)=0\), and \(P_{\mathrm{ASE},b}(L)=0\).

The signal absorption cross section was extracted from measurements of the small-signal loss in a 55-mm-long waveguide, whereas the pump absorption cross section was extracted from a 6.6-mm-long waveguide because of the stronger pump absorption. The Yb-related contribution to the loss was obtained by subtracting the coupling and passive waveguide losses from the measured small-signal loss, converted to the cross section using
\begin{equation}
\sigma_{12} = \frac{\alpha_{\mathrm{active}}}{\Gamma N_{\mathrm{Yb}}},
\end{equation}
where $\alpha_{\mathrm{active}}$ is the active absorption. At 976~nm, the emission and absorption cross sections were assumed to be equal, since previous work in Yb-doped alumina has shown that this wavelength is near the McCumber zero-line energy \citeSI{Agazzi:2012:Thesis}. Finally, the signal emission cross section was extracted from the measured saturated active gain of 6.2~dB/cm in a 6.6-mm-long waveguide. The gain is defined as
\begin{equation}
g = \Gamma_s\left[\sigma_{21,s}N_2(z)-\sigma_{12,s}N_1(z)\right]
\end{equation}
and since $N_1 = N_2 = N_{\mathrm{Yb}}/2$ at 50\% inversion, in the small-signal and high-pump regime, we obtain
\begin{equation}
\sigma_{21,s} = \sigma_{12,s} + \frac{2g}{\Gamma_s N_{\mathrm{Yb}}}.
\end{equation}
The resulting cross sections are $\sigma_{12,p} = 1.9\times10^{-20}~\mathrm{cm^2}$, $\sigma_{21,p} = 1.9\times10^{-20}~\mathrm{cm^2}$, $\sigma_{12,s} = 3.0\times10^{-22}~\mathrm{cm^2}$, and $\sigma_{21,s} = 5.2\times10^{-21}~\mathrm{cm^2}$. We can then calculate the saturation energy as $E_{\mathrm{sat}} = \frac{h \nu A_{\mathrm{eff}}}{\Gamma_s \left(\sigma_{21,s} + \sigma_{12,s}\right)} = 730~\mathrm{nJ}$ and the saturation power as $P_{\mathrm{sat},W} = \frac{E_{\mathrm{sat}}}{\tau_{21}} = 2.9~\mathrm{mW}$.

\newpage
\subsection{Fabrication}

\begin{figure}[H]
    \centering
    \includegraphics[width=\linewidth]{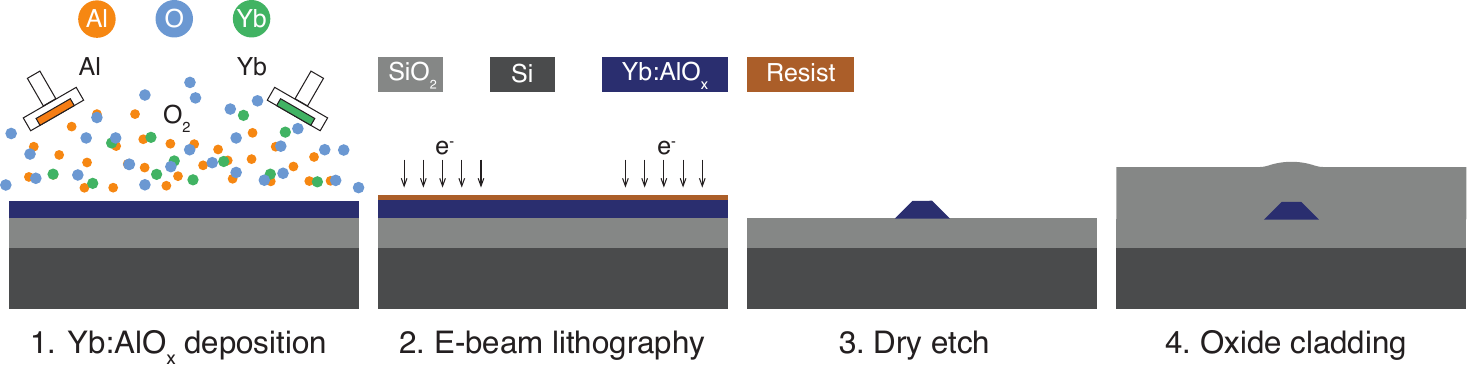}
    \caption{$\mathrm{Yb\!:\!AlO}_{\!x}$ photonic chip fabrication flow.}
    \label{fig:fabflow}
\end{figure}

\begin{figure}[H]
    \centering
    \includegraphics[width=\linewidth]{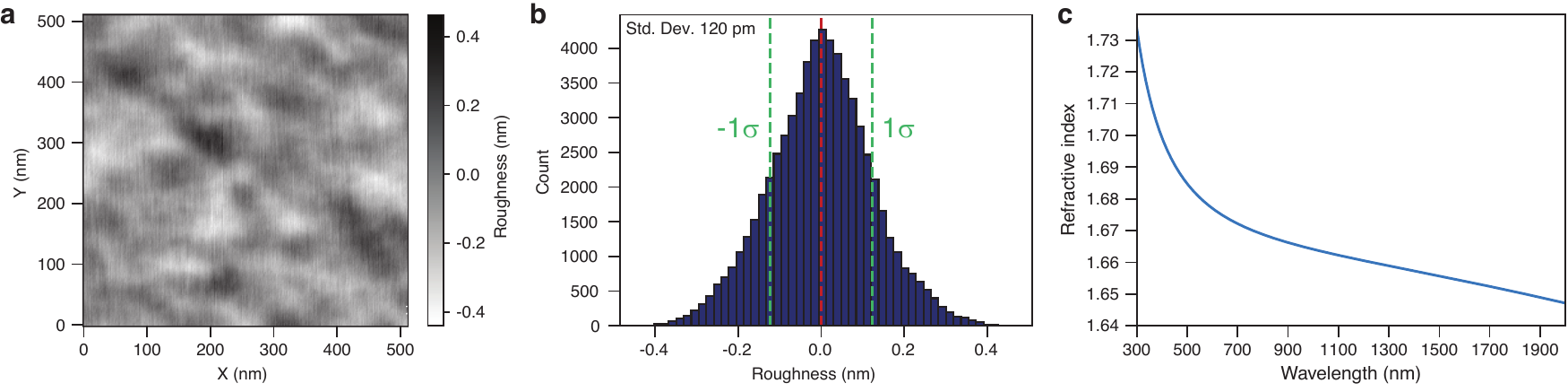}
    \caption{\textbf{Sputtered $\mathrm{Yb\!:\!AlO}_{\!x}$ film characteristics.}
    \textbf{(a)} Surface roughness measured by AFM.
    \textbf{(b)} Histogram of the AFM data showing a standard deviation of 120~pm.
    \textbf{(c)} Refractive index of the $\mathrm{Yb\!:\!AlO}_{\!x}$ thin film.
    } 
    \label{fig:AFM}
\end{figure}

The fabrication process of the ytterbium-doped AlO\textsubscript{x} photonic circuits started from the growth of the gain film (Fig. \ref{fig:fabflow}). 
We used an RF magnetron co-sputtering technique \citeSI{polman:2003:JAP,Pollnau:2009:IEEEJQE,yang:2025:CLEO:ErInjection} to deposit a 1.08-µm $\mathrm{Yb\!:\!AlO}_{\!x}$ film on top of a 4-µm wet thermal oxide layer, which served as the bottom cladding. The surface roughness of the sputtered film was measured by atomic force microscopy (AFM, Jupiter XR) to be 120~pm (Fig. \ref{fig:AFM}a-b). The thickness and refractive index of the $\mathrm{Yb\!:\!AlO}_{\!x}$ film were characterized by SEM and ellipsometry (Woollam), yielding a thickness of 1080~nm and a refractive index of 1.66 at 1030~nm (Fig. \ref{fig:AFM}c).

Nanophotonic structures were then patterned with electron-beam lithography (Elionix) on a negative resist (ma-N 2400 series) followed by a Cl-based reactive ion etching (RIE) process \citeSI{Pollnau:2007:APB}. 
The etch pressure was carefully tuned to produce smooth sidewalls with a sidewall angle of \ang{52} and a controllable lateral etch of 500–800~nm on each side of the waveguide. For waveguide patterns narrower than the total lateral etch offset, the etched structure evolves into a triangular cross section, with both height and width decreasing as the etch proceeds. Using this technique, we engineered a three-dimensional inverse taper at the amplifier facet (see Fig. 1  and Fig. \ref{fig:Chip_scheme}) to minimize facet reflection and insertion loss.

Following the patterning of the active region, 5~µm of $\mathrm{SiO_2}$ was deposited as the top cladding.  Finally, we annealed the devices at 725~\textdegree C for 4~hours to reduce the amount of hydrogen bonds in the PECVD oxide \citeSI{liu:2026:arxiv:KK}. The dies were then diced and polished for measurement.

The supercontinuum waveguides were fabricated in a similar fashion. Silicon nitride (SiN) was grown by low-pressure chemical vapor deposition (LPCVD) on a 4-µm wet thermal oxide layer. Waveguides were patterned in the same manner as described above, except that the RIE was performed using F-based chemistry. The SiN waveguides were annealed at 1050 \textdegree C for 8 hours after etching. We then deposited 2~µm of silicon oxide as the top cladding using LPCVD. Finally, we annealed the device again at 1050~\textdegree C for 8~hours.


\newpage

\subsection{CW amplification}
\begin{figure}[H]
    \centering
    \includegraphics[width=\linewidth]{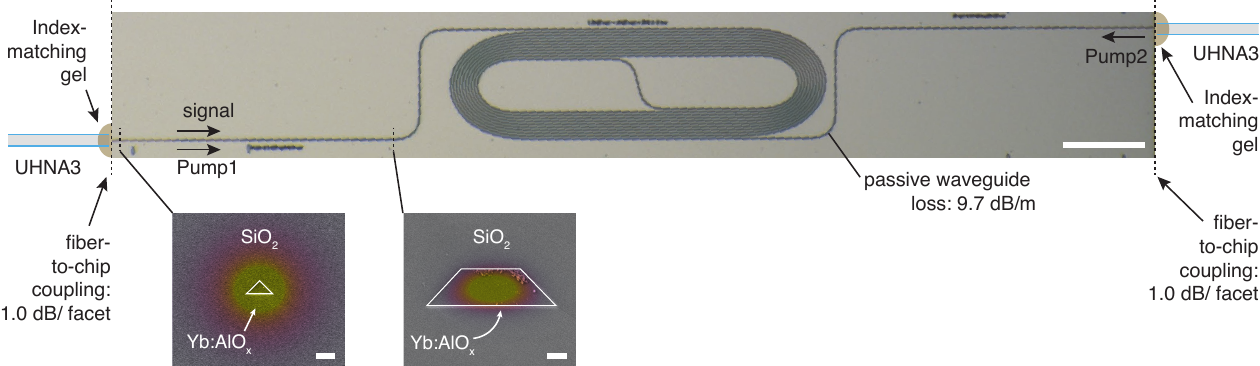}
    \caption{\textbf{Input and output fiber-coupled integrated Yb-doped amplifier.} The amplifier chip facets were polished close to normal to the input and output waveguide direction. The waveguides are tapered in both width and height to facilitate efficient coupling and low back-reflection. Cleaved UHNA3 fibers were butt-coupled to the waveguide alongside index-matching gel. SEM images of the waveguide cross-sections (also presented in Fig. 1 in the main text) are included for reference. The measured fiber-to-chip coupling loss and passive waveguide transmission losses are approximately 1.0 dB per facet and 9.7 dB per meter, respectively. }
    \label{fig:Chip_scheme}
\end{figure}

The amplifier was pumped bidirectionally using two 976 nm diode lasers (Thorlabs BL976-PAG900). The signal was provided by a tunable CW laser covering 1010–1100 nm (Toptica CTL). Two 980/1060 nm wavelength-division multiplexers (WDMs, Thorlabs WD9860AB) were used to combine the pump and signal at the input and to separate them at the output. Ultra-high-numerical-aperture fiber (Nufern/Coherent UHNA3) with a mode-field diameter of 2.6 µm at 1100 nm \citeSI{coherent:2020:uhna3} was used for chip coupling (Fig. \ref{fig:Chip_scheme}). All other fiber segments were HI 1060.

Input and output spectra were recorded with an optical spectrum analyzer (Yokogawa AQ6370D, 600--1700 nm) at a resolution bandwidth of 0.1 nm. The OSA was calibrated using an integrating-sphere power sensor (Thorlabs S145C). The calibrated measurements were then converted to off-chip and on-chip powers using the WDM and fiber-chip coupling loss. The optical gain was extracted by taking the difference between the peak output power and peak input power (Fig. \ref{fig:NF}).

\subsection{Ultrafast pulse amplification}\label{sec_pulse_amp}
The pulsed source used for the ultrafast pulse amplification is derived from an all-fiber Er-doped ultrafast frequency comb operating in the 1.55-µm spectral region. The oscillator generates stable femtosecond pulses that can be externally compressed to durations below 50 fs, enabling high peak power for efficient nonlinear conversion. These pulses are subsequently launched into a highly nonlinear fiber, where significant spectral broadening occurs via third-order nonlinear ($\chi^3$) processes. In particular, dispersive wave generation facilitates efficient energy transfer from the soliton to the 1-µm spectral region. The generated 1-µm dispersive wave component is spectrally filtered and used as a coherent seed for a Yb-doped fiber amplifier. After amplification, a stable 1-µm frequency comb is obtained with an average output power exceeding 150 mW and compressed pulse duration below 150 fs, while fully preserving comb coherence and phase stability.

The pulsed optical gain measurement used the same pumping scheme as the CW gain measurement.
A total of 4~m of HI 1060 fiber (including the fiber lengths constituting the polarization controller and the isolator) was inserted between the pulsed laser and the WDM to pre-stretch the pulse to a width of 5.2 ps. 
The same OSA resolution bandwidth of 0.1 nm was used for the measurement.
The pulse power was calculated from the measured spectra by accounting for the wavelength-dependent WDM loss to obtain the correct off-chip power and the on-chip pulse power was then calculated by compensating for the chip coupling loss of 1.0 dB per facet. 

After amplification, the pulses were compressed using an in-house pulse compressor consisting of two transmission gratings (Coherent LightSmyth), with a design wavelength of 1040 nm and a groove density of 1000 grooves/mm to provide anomalous dispersion. The output from the compressor was then sent to an autocorrelator (Femtochrome FR-103XL) for pulse width characterization.
\subsection{Supercontinuum generation}

\begin{figure}[H]
    \centering
    \includegraphics[width=1\linewidth]{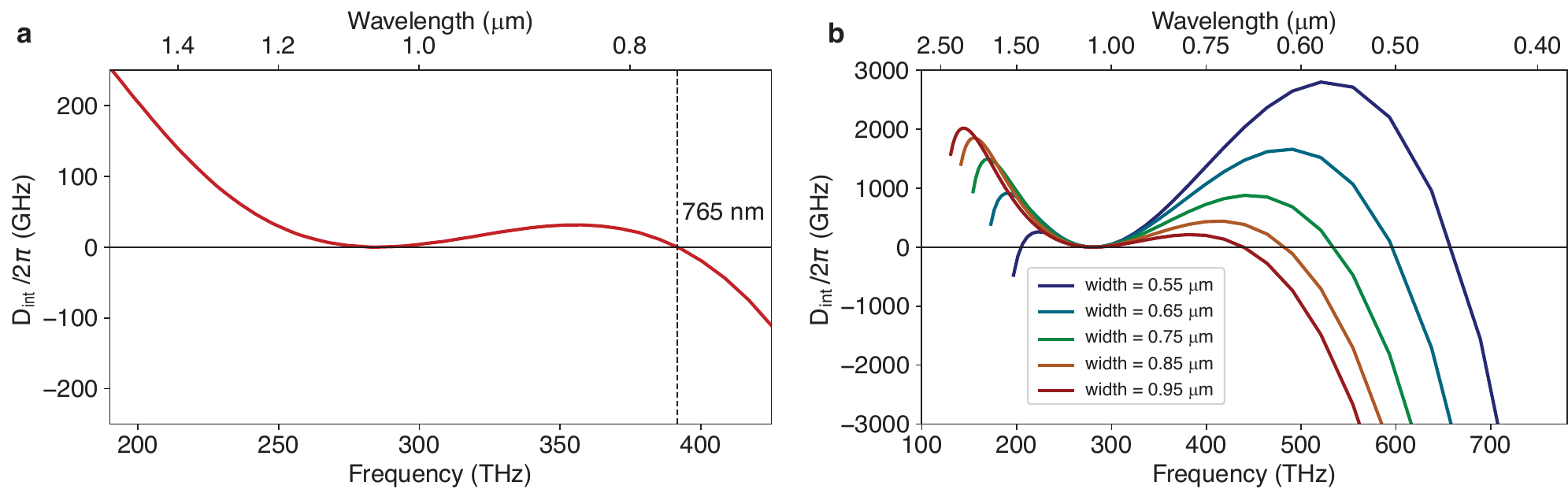}
    \caption{
    \textbf{Integrated dispersion for supercontinuum waveguides. (a)} Integrated dispersion for the TE mode in the $\mathrm{Si_3N_4}$ waveguides with a width of 0.9~$\mu$m. \textbf{(b)} Integrated dispersion of the TE mode in the $\mathrm{SiO_2\!:\!Ta_2O_5}$ waveguides with widths ranging from 0.55~$\mu$m to 0.95~$\mu$m.
    }
    \label{fig:dispersion}
\end{figure}

For the supercontinuum generation experiments, we adopted a pre-compression scheme to avoid compressor insertion loss. The remainder of the pulse amplification setup was identical to that described in section \ref{sec_pulse_amp}. Following amplification, the pulse was sent through an optical isolator to suppress reflections from the facets of the passive waveguides. Polarization was adjusted using a paddle-based fiber polarization controller, and lensed fibers were used for coupling into and out of the passive supercontinuum waveguides.

The output spectra were monitored using two OSAs to cover the wavelength range from 400 to 1800 nm: a short-wavelength OSA (350--1200~nm, Yokogawa AQ6373E) and a long-wavelength OSA (1200--2400~nm, Yokogawa AQ6375). In grating-based spectrum analyzers, shorter-wavelength light can overlap longer-wavelength signals through higher diffraction orders, producing spurious spectral features. The short-wavelength OSA suppresses these artifacts using built-in high order optical cut filters. Because the long-wavelength OSA does not include such filters, we inserted a long-pass filter with a cut-on wavelength of 1050~nm  to block the short-wavelength components. To compensate for the filter loss, the 1200–2400~nm spectrum was scaled to match the 350–1200~nm spectrum at the 1200~nm overlap point. 

We performed supercontinuum generation experiments in both $\mathrm{Si_3N_4}$ and $\mathrm{SiO_2\!:\!Ta_2O_5}$ waveguides. The $\mathrm{Si_3N_4}$ waveguides were fabricated in-house with a waveguide height of 700~nm, a width of 900~nm, and oxide cladding to target a dispersive wave around 765~nm. The integrated dispersion for the TE mode is shown in Fig. \ref{fig:dispersion}a.  The waveguide was 50~mm long and inverse-tapered at the facets to a width of 200~nm, improving the coupling efficiency to 6.3~dB/facet.
The $\mathrm{SiO_2\!:\!Ta_2O_5}$ waveguides with various waveguide widths were used to generate dispersive waves at visible wavelengths spanning from 476 to 644~nm \citeSI{papp:2020:OL,papp:2021:Optica}. 
To obtain the desired dispersion profile, the waveguides were air-clad and $\mathrm{SiO_2}$ was added as a dopant \citeSI{papp:2025:arxiv} (Fig. \ref{fig:dispersion}b). Waveguide tapers were implemented to improve the coupling efficiency to 3.7~dB/facet for the $\mathrm{SiO_2\!:\!Ta_2O_5}$ waveguides.

The phase coherence of the $\mathrm{Si_3N_4}$ supercontinuum generation was evaluated by measuring the repetition  rate beatnotes across the generated spectrum by using different combinations of optical bandpass filters and photodetectors.
One photodetector covers the wavelength region of 900--1700~nm (Newport 1811-FC-AC, PD1), while the other covers a wavelength range of 320--1000~nm (Newport 1801-FC-AC, PD2). The two bandpass filters cover the wavelength ranges of 1045--1055~nm (Thorlabs FBH1050-10, BP1) and 778.5--781.5~nm (Thorlabs FBH780-3, BP2).

\clearpage
\section{Additional information}
\vspace{-14pt}
\subsection{Passive loss characterization of ytterbium-doped alumina waveguides} \label{sec_loss}
\begin{figure}[H]
    \centering
    \includegraphics[width=0.6\linewidth]{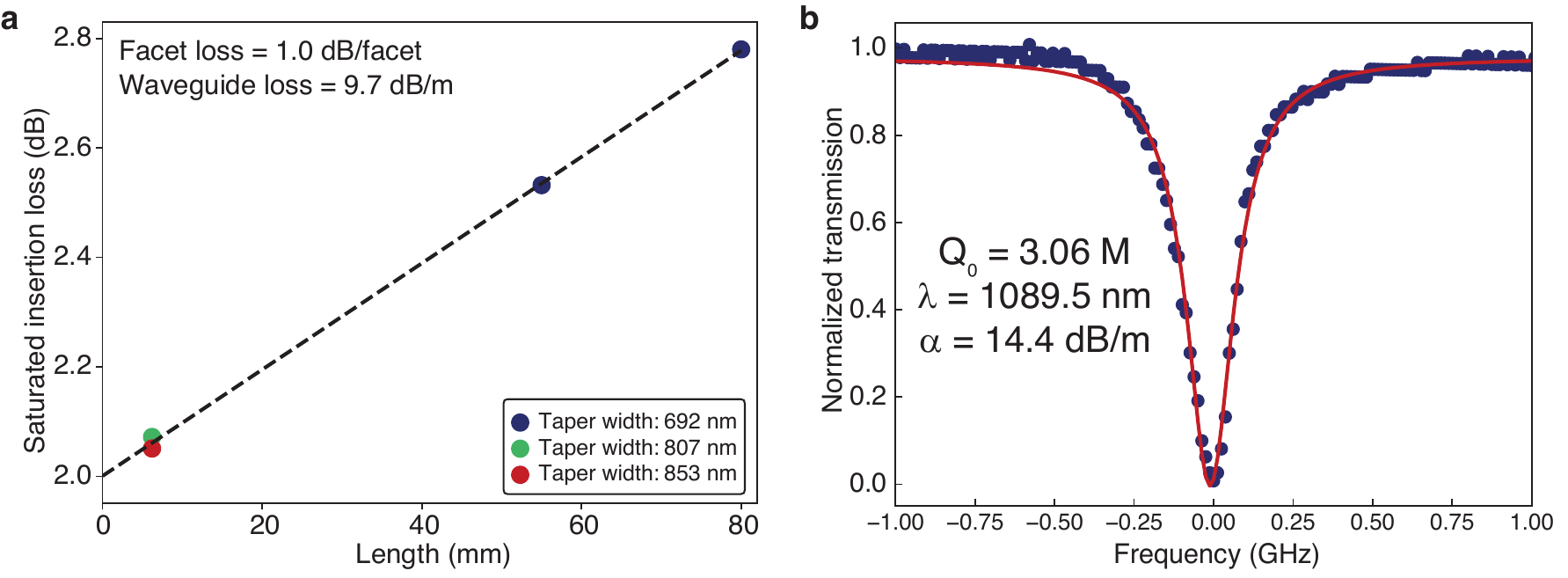}
    \caption{\textbf{Passive insertion loss characterization.} \textbf{(a)} Cutback measurement for passive waveguide loss and coupling loss extraction. \textbf{(b)} Microring resonator quality factor measurement to extract the passive waveguide loss. $\mathrm{Q_0}$, fitted intrinsic quality factor; $\lambda$, center wavelength; $\alpha$, inferred propagation loss.}
    \label{fig:Passive_loss}
\end{figure}

\vspace{-14pt}
The passive insertion loss of the amplifier was measured at 1060 nm by using a YDFA to drive the device into gain saturation, thereby isolating the passive loss. A wavelength of 1060 nm was used instead of 1030 nm due to the operation bandwidth of the YDFA. To separate coupling loss from passive waveguide loss, we measured both 80-mm spiral amplifiers, 55-mm spiral amplifiers, and short 6.6-mm amplifiers. Because of yield limitations, the short devices could not be measured with the same 692-nm bottom taper width, so devices with 807-nm and 853-nm taper widths were characterized instead. These showed identical insertion losses of 2.1 dB. Given this close agreement, data from all taper widths were fit together, yielding an estimated coupling loss of 1.0 dB per facet and a passive waveguide loss of 9.7 dB/m (Fig. \ref{fig:Passive_loss}a). To validate this result, we also measured the quality factors of 100-\textmu m-radius microring resonators from a subsequent fabrication batch at 1089.5 nm to avoid the Yb absorption at 1060 nm, which gave an inferred loss of 14.4 dB/m (Fig. \ref{fig:Passive_loss}b). The input power for the quality factor measurements was increased to just below the onset of the thermal bistability, so that the active loss was maximally saturated while avoiding thermally induced resonance shifts that would complicate quality factor extraction. The measured loss is higher than the cutback value, which we attribute to residual active loss and bending loss.
\vspace{-10pt}

\begin{figure}[H]
    \centering
    \includegraphics[width=0.6\linewidth]{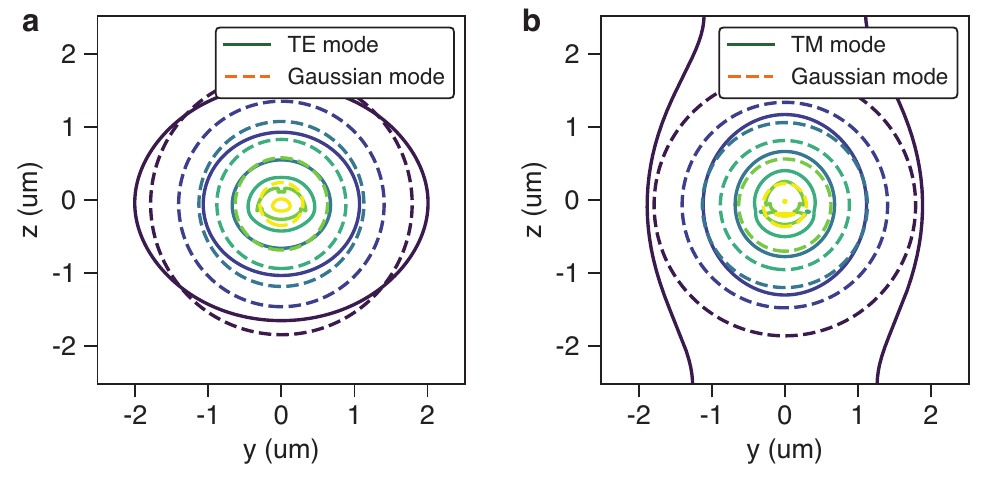}
    \caption{
    \textbf{Optical mode profiles in the 3D inverse taper and UHNA3 fiber.}
    \textbf{(a)} Mode overlap between the 1030-nm TE mode of a 3D inverse-taper and that of UHNA3 fiber, with 
    an overlap factor of 92.0\%. A Gaussian mode 
    with a mode field diameter of 2.6 µm was used 
    to represent the mode profile in the UHNA3 fiber.
    \textbf{(b)} Mode overlap between the 1030-nm TM mode of a 3D inverse-taper and that of UHNA3 fiber, with 
    an overlap factor of 87.9\%. 
    }
    \label{fig:overlap}
\end{figure}

The simulated mode profile of the 3D inverse-taper and the UHNA3 fiber Gaussian mode is shown in Fig. \ref{fig:overlap}, resulting in a mode overlap of 92.0\% for waveguide TE mode. The mode overlap factor is calculated using the following equation:

{\setlength{\abovedisplayskip}{2pt}
\setlength{\belowdisplayskip}{2pt}
\setlength{\abovedisplayshortskip}{2pt}
\setlength{\belowdisplayshortskip}{2pt}
\begin{equation}
\eta
=
\left|
\frac{
\iint
\left[
E_{1y}^{*}(y,z)E_{2y}(y,z)
+
E_{1z}^{*}(y,z)E_{2z}(y,z)
\right]
\,dy\,dz
}{
\sqrt{
\iint
\left[
\left|E_{1y}(y,z)\right|^2
+
\left|E_{1z}(y,z)\right|^2
\right]
\,dy\,dz
}
\sqrt{
\iint
\left[
\left|E_{2y}(y,z)\right|^2
+
\left|E_{2z}(y,z)\right|^2
\right]
\,dy\,dz
}
}
\right|^2 .
\end{equation}
}

\subsection{Rutherford backscattering spectrometry} 
\label{sec_RBS}

\begin{figure}[H]
  \centering
  \includegraphics[width=1\linewidth]{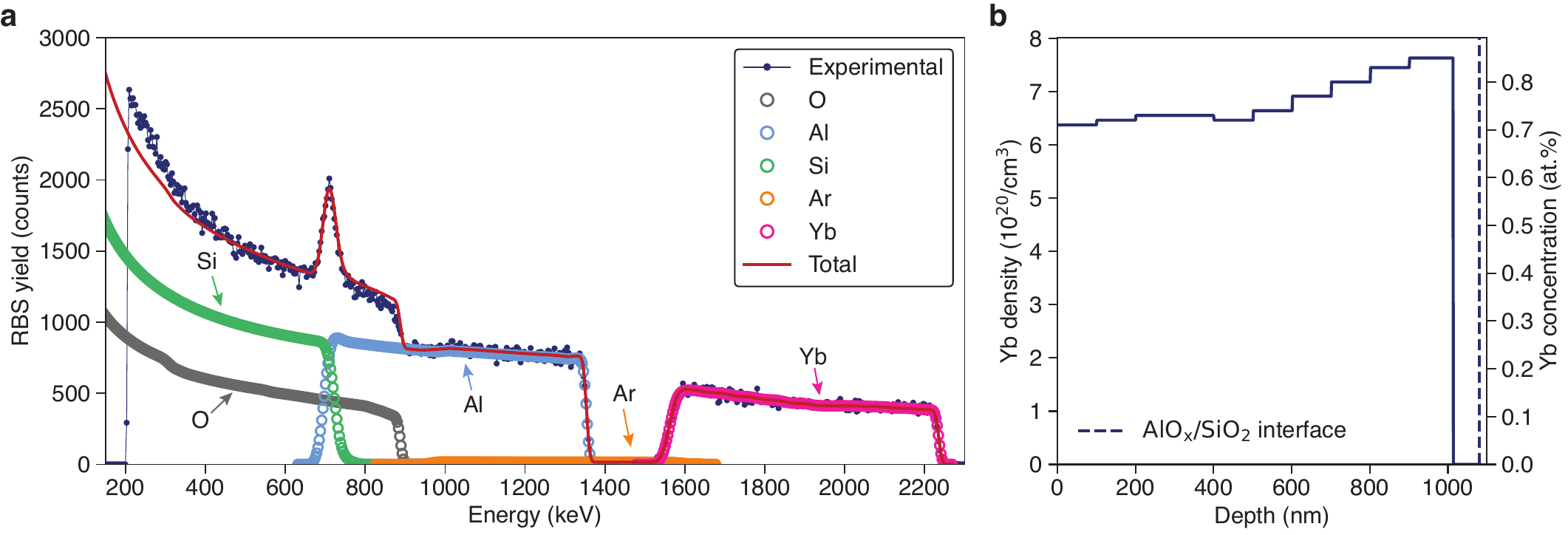}
  \caption {\textbf{Rutherford backscattering spectrometry.} \textbf{(a)} Measured and simulated RBS spectra, individual element contributions are shown as separate traces. \textbf{(b)} Yb concentration depth profile.}
  \label{fig:RBS}
\end{figure}

We employ Rutherford backscattering spectrometry (RBS) to determine the film stoichiometry and Yb doping concentration. For the measurement, 2.45 MeV He$^+$ ions were used with a scattering angle of 170$^\circ$. Simulations and fitting of the RBS spectrum were performed using SIMNRA by the authors. The measured spectrum, fitting, and calculated Yb density are shown in Fig. \ref{fig:RBS}. In order to obtain an accurate profile of various elements over the entirety of the film thickness, a total of 11 layers were used to fit the RBS spectrum. One layer was used to represent the undoped $\mathrm{AlO_x}$ layer at the beginning of the growth (deepest), while the remaining 10 layers of similar thickness were used to represent the doped $\mathrm{AlO_x}$ region. The material composition and thickness of all layers were fine-tuned to achieve the best fit. A thick layer of stoichiometric $\mathrm{SiO_2}$ was used underneath for the simulation, which matches the lower energy spectrum decently well. There is a trace amount of Ar within the film based on the simulation as expected, since Ar is the dominant sputtering gas used during the deposition process.

Between the energy of 1550 keV and 2250 keV, the spectrum is dominated by the Yb scattering signal since it is the heaviest element in the film. With a large atomic number, Yb also has a large scattering cross-section. Consequently, even at concentrations as low as 0.7-0.8 at.\% (atomic percentage), the He$^+$ counts scattered from Yb are comparable to those scattered from the Al and O. The high Yb sensitivity enables a highly accurate fit of the Yb concentration at different depths, as shown in Fig. \ref{fig:RBS}b. 
At the low-energy edge near 1550 keV, the measured profile can be accurately reproduced by tuning the Yb concentration in the deepest layer and the Ar concentration near the surface. The best agreement is obtained for a gradual Ar profile near the surface, in which the atomic fraction increases from 0.1 at.\% at the surface to 0.4 at.\% in the deeper region.
The location of this edge reflects the total areal atom density, since He$^+$ ions lose energy as they undergo scattering along their trajectory through the film. Accordingly, the minimum detected energy of He$^+$ ions scattered from Yb is determined by the total areal atomic density encountered by the ion beam, expressed in  $\mathrm{atom/cm^2}$.
This effect allows us to fit the total areal atom density of the film accurately.

After determining the Yb and Ar concentrations, the Al concentration in the top two thirds of the alumina film, corresponding to the 900–1400 keV region, is obtained by fitting the RBS signal. The O concentration in this upper portion of the film is then taken as the remaining atomic fraction not accounted for by the other elements.
Extracting the Al concentration in the bottom third of the alumina film, corresponding to 700–900 keV, is more challenging because its signal overlaps with the oxygen contribution from the upper portion of the film. However, because the oxygen concentration in the top two thirds has already been determined, this overlapping contribution can be accounted for, allowing the Al concentration in the bottom third to be fitted. The O concentration in the bottom third is then again taken as the remaining atomic fraction not accounted for by the other elements. The total density previously determined from the Yb trace is then verified using the energy span of the Al scattering signal, which produces the sharp peak at 700 keV together with Si.
We note that the minor discrepancies at low energies (below 400 keV) are a result of uncertainties in the stopping powers, which describe the rate of energy loss for He$^+$ ions as they travel through the material. These discrepancies do not affect the accuracy of our fitting, since none of the Al, Yb, and Ar signals appear in this range.

Following this high-to-low energy, element-by-element fitting procedure, we precisely determined the stoichiometry and total areal density of the $\mathrm{AlO_x}$ film. The O atomic fraction ranges between 61 at.$\%$ and 62 at.$\%$, with an average of 61.5 at.$\%$. The Al atomic fraction ranges from 36.8 at.$\%$ to 38.2 at.$\%$, with an average of 37.4 at.$\%$. This means that the alumina $(\mathrm{Al_2O_{3.28}}$) film is oxygen-rich. The average Yb atomic fraction within the doped region is 0.76 at.$\%$, corresponding to an average density of $\mathrm{6.8\times10^{20}\: atoms/cm^3}$. The total number of atoms per unit area obtained from the fitting is $\mathrm{9.70\times 10^{18}\:at./cm^2}$. Combining the molar masses of Al, O, and Yb with the 1080~nm film thickness, we can calculate an average film density of 3.17 $\mathrm{g/cm^3}$ for the doped region and 2.99 $\mathrm{g/cm^3}$ for the undoped region. 
The Yb-related He$^+$ ion counts have a standard deviation of 19 counts, corresponding to a concentration uncertainty of 0.034 at.\%. We therefore report the Yb concentration to a resolution of 0.01 at.\%.

\noindent 
\\

\subsection{Photoluminescence lifetime measurement} 
\label{sec_Lifetime}

\begin{figure}[hbt!]
    \centering
    \includegraphics[width=0.4\linewidth]{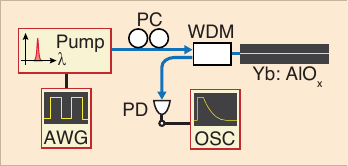}
    \caption{Experimental setup for photoluminescence lifetime measurements.}
    \label{fig:Lifetime}
\end{figure}
The photoluminescence lifetime was measured by modulating a 976-nm pump coupled into a 6.6-mm-long waveguide. The resulting amplified spontaneous emission (ASE) was collected from the 1060-nm port of a WDM and detected with a Newport 1811 InGaAs photoreceiver. The electrical signal was recorded on a Keysight DSOX3014G oscilloscope in high-resolution mode to reduce noise, then fit with a single-exponential decay to extract the Yb lifetime. A low pump power of 3.7 mW was used to minimize quenching-induced lifetime shortening. The 976-nm pump was modulated at 10 Hz using the Thorlabs ITC4005 controller, with the modulation synchronized to the oscilloscope trigger. 

\newpage
\subsection{DFB laser characterization} 

\begin{figure}[hbt!]
    \centering
    \includegraphics[width=0.85\linewidth]{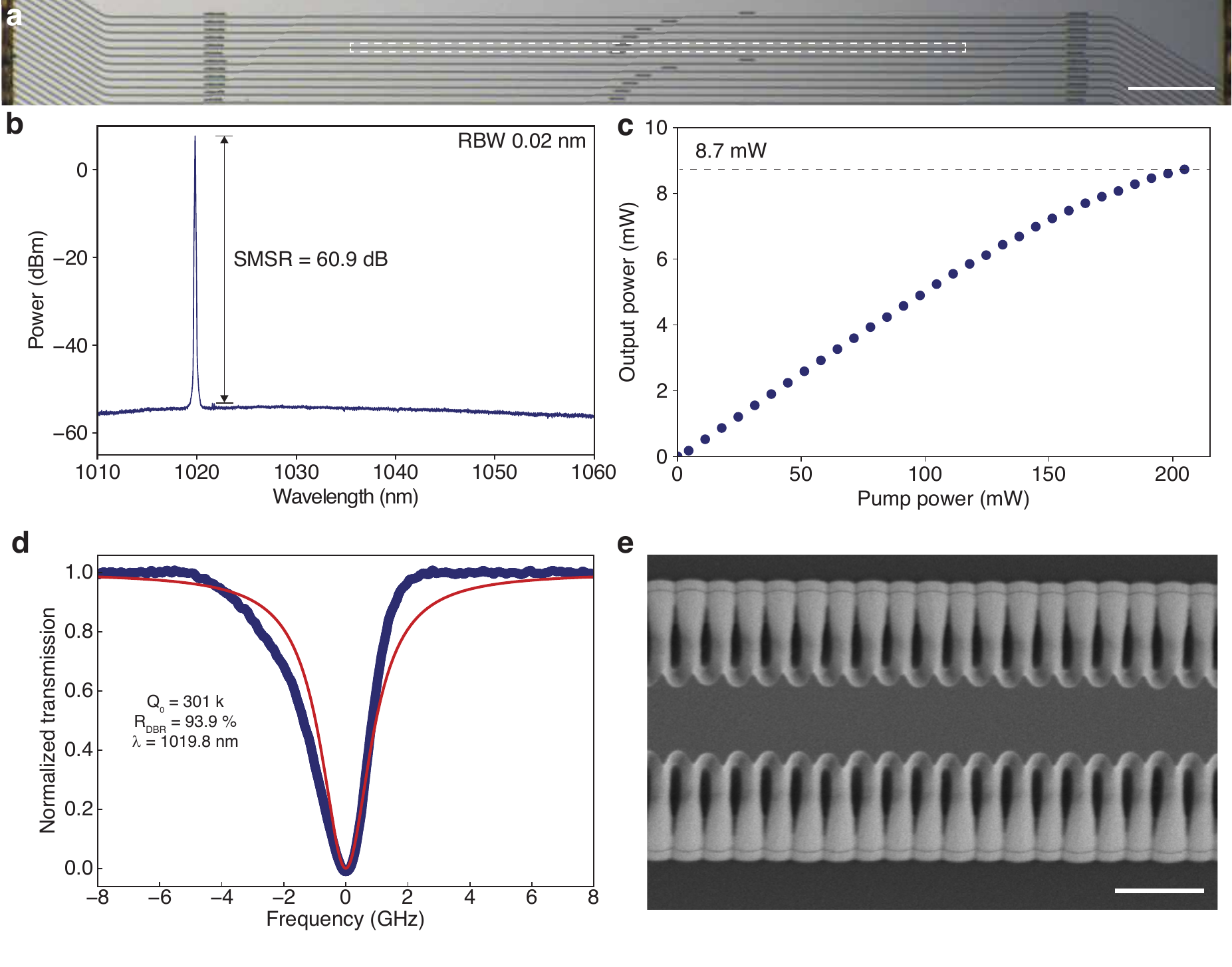}
    \caption{\textbf{DFB laser characterization.} 
    \textbf{(a)} Optical image of a DFB array. The DFB under test is highlighted. The scale bar is 500 µm.
    \textbf{(b)} DFB laser spectrum showing a side mode suppression ratio (SMSR) of 60.9 dB. A resolution bandwidth (RBW) of 
    0.02 nm was used for the measurement.
    \textbf{(c)} Measured output power at 1019.8 nm.
    \textbf{(d)} Measured and fitted reflection spectra of the DFB laser cavity showing a Bragg mirror reflectivity of 93.9\%. 
    \textbf{(e)} SEM image of a section of the Bragg mirror. The scale bar is 1 µm.
    }
    \label{fig:DFB}
\end{figure}

We demonstrate a DFB laser formed by two 1.6-mm-long Bragg gratings, each with a width of 1.4 \textmu m (before lateral etch offset) and 100 nm peak-to-peak sidewall corrugations, separated by a 60 \textmu m cavity. When pumped uni-directionally at 976 nm, the device produces a maximum on-chip output power of 8.7 mW at 1019.8 nm and exhibits a side-mode suppression ratio as high as 60.9 dB.
The output was collected with 1064 nm lensed fibers (TSMJ-3A-1064-6/125/-0.25-10-2-10-1) and characterized through a 980/1060 nm fiber wavelength-division multiplexer using a Yokogawa AQ6375 optical spectrum analyzer with a 0.02-nm resolution bandwidth. To suppress facet reflections, the output waveguides were angled by \ang{32}.

\subsection{Optical gain and noise figure measurement}

\begin{figure}[H]
  \centering
  \includegraphics[width=0.45\linewidth]{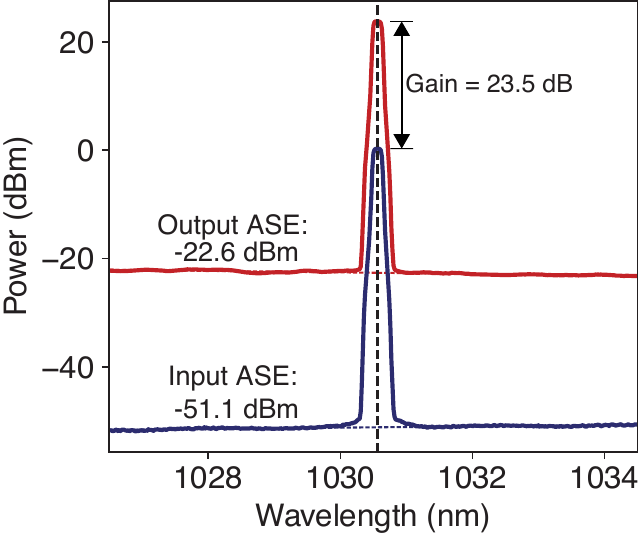}
  \caption{Optical spectra of the on-chip input and amplified output signals used to calculate  the lowest noise figure of 3.3 dB.}
  \label{fig:NF}
\end{figure}

The noise figure (NF) was calculated using the input and output laser spectra acquired by an optical spectrum analyzer (Yokogawa AQ6370D). Fig. \ref{fig:NF} shows the input and output spectra for the lowest noise figure of 3.3 dB. For the NF calculations, we used the following definition \citeSI{Haus:1998:IEEE} which includes the 1/G shot noise and subtracts the input laser's source spontaneous emission (SSE):
\begin{equation}
F = \frac{P_{\mathrm{ASE,out}}-GP_{\mathrm{SSE}}}{G\,h\nu B_0} + \frac{1}{G}.
\end{equation}
Here, $P_{\mathrm{ASE,out}}$ and $P_{\mathrm{SSE}}$ are the output ASE and input SSE powers respectively, 
G is the optical gain, h is Planck's constant, $\nu$ is the peak frequency, $\mathrm{B_0}$ is the effective noise bandwidth, and F denotes the noise figure.
To ensure accurate power scaling, the OSA bin power was normalized with the Thorlabs S145C power monitor and the effective noise bandwidth was calibrated to a value of 0.144 nm. We treat the measured ASE power as the combined ASE contribution from both polarizations due to the similar fiber-mode overlap factors (section \ref{sec_loss}).  
The minimum measured on-chip noise figure of 3.3 dB is enabled by the low spontaneous emission factor ({$n_{sp}$}), which is given by

\begin{equation}
\begin{gathered}
\begin{aligned}
\eta = \sigma_{21,s}/\sigma_{12,s} , \\
n_{sp} =\frac{ \eta N_2}{\eta N_2-N_1}, \\
\mathrm{NF}_{min} = 2n_{sp}.
\end{aligned}
\end{gathered}
\label{eq:rate}
\end{equation}
Setting {$N_2=N_1$} due to equal pump emission and absorption cross sections for the pump yields a minimum noise figure of 3.29 dB using the cross sections obtained in section \ref{sec_modeling}. An {$n_{sp}$} value of 1 corresponds to the quantum limited 3 dB noise figure.

\label{sec_NF}

In this paper, we assume equal coupling losses at the input and output facets of the amplifier. However, at a pump power of 52.7 mW, the ASE measured from the input side was about 2.5 dB higher than that measured from the output side. This suggests that the input coupling loss is lower than the assumed 1.0 dB per facet. As a result, the actual noise figure is likely higher than the calculated value of 3.3 dB assuming symmetric facet coupling loss.

% \clearpage
\subsection{Device performance comparison}

Table \ref{tab:CW_amp} shows a comparison of CW amplifiers with integrated rare-earth gain, transition-metal gain, rare-earth fibers, and III-V amplifiers. 
Table \ref{tab:Pulsed_Amp} shows a comparison of pulse amplifiers. The 730~nJ saturation energy of the $\mathrm{Yb\!:\!AlO}_{\!x}$ gain platform is advantageous compared to III-V SOAs with lower saturation energies of 5--10 pJ \citeSI{Agrawal:1989:IEEE_quantum_electronics} and shorter carrier lifetimes on the order of 100 ps \citeSI{Agrawal:1989:IEEE_quantum_electronics}. For pulse energies above the saturation energy, the beginning of the pulse will experience higher gain than the tail of the pulse. Additionally, the change in carrier populations creates an index shift, resulting in an effective n$_2$ coefficient on the order of $n_2\!= 1\!\times\!10^{-13}\ \mathrm{m^2/W}$ \citeSI{Agrawal:1989:IEEE_quantum_electronics}, which is around 5--6 orders of magnitude larger than the alumina platform. 

%\newpage
\renewcommand{\thetable}{S\arabic{table}}

\begin{table*}[p]
\centering
\scriptsize
\setlength{\tabcolsep}{3pt}
\renewcommand{\arraystretch}{1.15}

\caption{Integrated CW Amplification}
\label{tab:CW_amp}

\begin{tabular*}{\textwidth}{@{\extracolsep{\fill}} l c c c c c c l}
\toprule
\textbf{Active Medium} &
\textbf{WL} &
\textbf{Output} &
\textbf{Eff.} &
\textbf{On-chip} &
\textbf{Off-chip} &
\textbf{NF} &
\textbf{Reference} \\
&
\textbf{(nm)} &
\textbf{(W)} &
\textbf{(\%)} &
\textbf{gain (dB)} &
\textbf{gain (dB)} &
\textbf{(dB)} &
\\
\midrule

$\mathrm{Yb\!:\!AlO}_{\!x}$ &
1030 &
0.590 &
71 &
28.6 &
26.6 &
3.3 &
This work \\
\hline
Ti:Sapphire &
790 &
$\sim$0.06 &
12.3 &
$\sim$25 &
- &
- &
\citeSI{Vuckovic:2024:Nature} \\

Yb:AlO$_x$ &
1020 &
- &
- &
2.5 &
- &
- &
\citeSI{Pollnau:2025:Opt_mat} \\

Yb:LNOI &
1060 &
- &
5 &
8 &
-10 &
- &
\citeSI{Ya:2022:MDPI} \\

Yb:LNOI &
1064 &
7.94$\times\!10^{-9}$ &
- &
9 &
-1.72 &
- &
\citeSI{Xu:2023:Optics_Letters} \\

Nd:Al$_2$O$_3$ &
1064 &
2.2$\times\!10^{-3}$ &
1.4 &
19.9 &
14.5 &
8.5 &
\citeSI{blanco:2026:OE:NdAlO} \\

Nd:Al$_2$O$_3$ &
1064 &
2.75$\times\!10^{-4}$ &
- &
14.4 &
- &
- &
\citeSI{pollnau:2010:APB:Nd} \\

Nd:Al$_2$O$_3$ &
1330 &
3.23$\times\!10^{-5}$ &
- &
5.1 &
- &
- &
\citeSI{pollnau:2010:APB:Nd} \\

Er:Si\textsubscript{3}N\textsubscript{4}\textsuperscript{a} &
1550 &
0.145 &
60 &
30 &
24 &
7 &
\citeSI{kippenberg:2022:Science} \\

Er:Si\textsubscript{3}N\textsubscript{4} &
1550 &
0.022 &
12 &
15 &
5.2 &
- &
\citeSI{kippenberg:2024:OFC:MultiLane} \\

Er:Si\textsubscript{3}N\textsubscript{4} &
1560 &
0.055 &
- &
29.8 &
24 &
- &
\citeSI{kippenberg:2024:arxiv:terabit} \\

Er:Si\textsubscript{3}N\textsubscript{4} &
1550 &
2.3$\times\!10^{-3}$ &
- &
- &
12 &
- &
\citeSI{kippenberg:2025:CLEO:pol} \\

Er:Si\textsubscript{3}N\textsubscript{4} &
1560 &
0.241 &
28.4 &
16 &
13.5 &
- &
\citeSI{kippenberg:2025:CLEO:highpower} \\

Er:Yb:LNOI\textsuperscript{b} &
1531 &
1.2$\times\!10^{-7}$ &
- &
8.26 &
- &
- &
\citeSI{xu:2023:PR:ErYbLN} \\

Er:Yb:LNOI &
1532 &
4.07$\times\!10^{-4}$ &
- &
27 &
10 &
- &
\citeSI{cheng:2023:OL:ErYbLN} \\

Er:LNOI &
1532 &
0.036 &
- &
38 &
24.6 &
- &
\citeSI{wang:2024:CLEO:ErLN:38dB},\citeSI{wang:2025:NatComm:ErLN:EO} \\

Er:LNOI &
1533 &
0.063 &
- &
35 &
- &
- &
\citeSI{li:2025:CLEO:ErLN:LMA} \\

Er:LNOI &
1531 &
4.0$\times\!10^{-5}$ &
- &
6 &
- &
4.2 &
\citeSI{zou:2026:PR:ErLN:modulator} \\

Er:Ta$_2$O$_5$ &
1532 &
- &
- &
4.8 &
- &
- &
\citeSI{wilkinson:2012:JLT:ErTantala:amp},\citeSI{wilkinson:2012:OC:ErTantala} \\

Er:Ga$_2$O$_3$ &
1532 &
1.7$\times\!10^{-7}$ &
- &
4.7 &
- &
- &
\citeSI{wang:2023:APL:ErGaO} \\

Er:Al$_2$O$_3$\textsuperscript{c} &
1533 &
1.3$\times\!10^{-7}$ &
- &
6.27 &
-1.23 &
- &
\citeSI{li:2024:LPR:GSS:ErAlO} \\

Er:Al$_2$O$_3$\textsuperscript{d} &
1532 &
2.51$\times\!10^{-3}$ &
- &
20 &
0 &
- &
\citeSI{blanco:pollnau:2014:OE:ErAlO} \\

Er:Al$_2$O$_3$ &
1556 &
0.088 &
16.3 &
28.5 &
24.3 &
4.6 &
\citeSI{blanco:2025:OE:broadband} \\

Er:Al$_2$O$_3$ &
1531 &
0.021 &
- &
33.5 &
-16.5 &
- &
\citeSI{blanco:2024:OE:ErAlO} \\

Er:glass &
1536 &
2.8$\times\!10^{-7}$ &
- &
6.5 &
4.5 &
- &
\citeSI{becker:2002:IEEE:ErGlass} \\

Er:Al$_2$O$_3$ on SiN &
1533 &
7.9$\times\!10^{-7}$ &
- &
1.98 &
- &
- &
\citeSI{sun:2019:NC:ALDErSiN} \\

Er:Al$_2$O$_3$ on SiN &
1532 &
6.3$\times\!10^{-5}$ &
- &
18.1 &
-1.9 &
- &
\citeSI{blanco:2020:PhotonicsResearch:ErAlO:SiN} \\

Er:Al$_2$O$_3$ on SiN &
1560 &
23.5$\times\!10^{-3}$ &
- &
16.33 &
13.73 &
$\sim$3.1 &
\citeSI{blanco:2025:OE:ErAlO:SiN} \\

Er:Yb:Al$_2$O$_3$ &
1531 &
1.41$\times\!10^{-4}$ &
- &
4.3 &
-4.7 &
- &
\citeSI{bradley:2020:OE:codope:solubility:wetetch} \\

Er:TeO$_2$ &
1530 &
0.02 &
- &
14 &
3 &
- &
\citeSI{madden:2010:OE:ErTeO2} \\

Er:TeO$_2$ on SiN &
1558 &
3.2$\times\!10^{-5}$ &
- &
5 &
- &
- &
\citeSI{bradley:2020:PR:ErTeO2SiN} \\

Tm:TeO$_2$ on SiN &
1870 &
7.6$\times\!10^{-5}$ &
- &
7.8 &
-4.2 &
- &
\citeSI{bradley:2019:OL:TmTeO2:onSiN} \\

Tm:Al$_2$O$_3$ on SiN &
1850 &
$\sim$1--1.8 &
63--66 &
16.5 &
11.2 &
5 &
\citeSI{Kaertner:2025:Nat_phot},\citeSI{Kaertner:2025:LSA} \\

Tm:Al$_2$O$_3$ on SiN &
1879 &
0.815 &
43 &
30 &
22 &
3.7 &
\citeSI{kartner:2025:arxiv:LMAamp} \\

\midrule

Yb:fiber &
- &
4.8 &
75 &
- &
- &
- &
Thorlabs\textsuperscript{e} \\

Yb:fiber &
1050 &
0.35 &
- &
- &
29 &
5.3 &
Thorlabs\textsuperscript{f} \\

\midrule

III-V MQW on Si\textsuperscript{g} &
1570 &
0.013 &
- &
13 &
3 &
5 &
\citeSI{bowers:2007:OFC:intel},\citeSI{bowers:2007:IEEE:intel} \\

III-V MQW on Si &
1550 &
0.048 &
- &
25.5 &
7.5 &
- &
\citeSI{bowers:2016:IEEE:QW:bondSi} \\

III-V MQW on Si &
1575 &
0.056 &
- &
27 &
8 &
$\sim$9 &
\citeSI{Roelkens:2019:OE} \\

III-V MQW on Si &
2001 &
3.16$\times\!10^{-5}$ &
- &
12 &
- &
- &
\citeSI{bowers:2016:CLEO:2umSOA} \\

III-V MQW &
1560 &
4$\times\!10^{-3}$ &
- &
20 &
- &
- &
\citeSI{lealman:2008:OC:SOA:QuantumWell:C-band} \\

III-V QD\textsuperscript{h} &
1300 &
0.01 &
- &
16 &
- &
- &
\citeSI{fujitsu:2003:PSS:patter:XGM:free} \\

III-V QD &
1300 &
8$\times\!10^{-3}$ &
- &
18 &
-6 &
8 &
\citeSI{tunnermann:2003:IEEEJQE:1300QD} \\

III-V QD on Si\textsuperscript{i} &
1315 &
0.229 &
- &
39 &
26.9 &
8.5 &
\citeSI{bowers:2019:ACS:Oband:QD:grown} \\

III-V QD &
1305 &
0.087 &
- &
18.2 &
- &
5.6 &
\citeSI{plant:2025:JLT:ObandComp:innolume} \\

III-V QD &
1450 &
0.2 &
- &
25 &
- &
5 &
\citeSI{arakawa:2005:IEEE:cbandSOA} \\

III-V QD &
1450 &
0.016 &
- &
18 &
- &
5.6 &
\citeSI{arakawa:2010:OFC:Cband:Qdot} \\

III-V QD &
1510 &
0.052 &
- &
37 &
22.5 &
9 &
\citeSI{jang:2007:APL:1510nm:37dB} \\

III-V TSA\textsuperscript{j} &
780 &
0.5 &
- &
- &
20 &
- &
\citeSI{Ye:2006:Optics_letters} \\

III-V MQW $\mu$TP SiN\textsuperscript{k} &
1570 &
7.6$\times\!10^{-3}$ &
- &
14 &
- &
10.6 &
\citeSI{roelkens:2020:Optica:uTPonSiN} \\

III-V $\mu$TP on Si &
1550 &
3.1$\times10^{-4}$ &
- &
8.1 &
- &
- &
\citeSI{roelkens:2025:SPIE:uTPSOA} \\

III-V MQW $\mu$TP on Si &
1573 &
0.079 &
- &
9 &
- &
- &
\citeSI{roelkens:2023:IEEE:uTP} \\

BOA\textsuperscript{l} &
1040 &
$>69\!\times\!10^{-3}$ &
- &
- &
28 &
7.5 &
Thorlabs\textsuperscript{m} \\

\bottomrule
\end{tabular*}

\vspace{0.5em}
\begin{minipage}{\textwidth}
\footnotesize
\textsuperscript{a} Er implanted in Si\textsubscript{3}N\textsubscript{4} waveguides.
\textsuperscript{b} LNOI: lithium niobate on insulator.
\textsuperscript{c} Ge$_{25}$Sb$_{10}$S$_{65}$ waveguides on Er:Al$_2$O$_3$ deposited by atomic layer deposition (ALD).
\textsuperscript{d} Fully-etched Er:Al$_2$O$_3$ amplifier waveguides heterogeneously integrated with Si$_3$N$_4$ photonic circuits.
\textsuperscript{e} Model number: YB1200-6/125DC.
\textsuperscript{f} Model number: YDFA300P.
\textsuperscript{g} MQW: multi-quantum well. III-V bonded on Si unless otherwise specified.
\textsuperscript{h} QD: quantum dot.
\textsuperscript{i} Directly epitaxially grown on Si.
\textsuperscript{j} TSA: tapered semiconductor amplifier.
\textsuperscript{k} $\mu$TP: micro-transfer print.
\textsuperscript{l} BOA: booster optical amplifier.
\textsuperscript{m} Model number: BOA1050S.
\end{minipage}

\end{table*}
%%%%%%%%%%%
%%% Table 2
%%%%%%%%%%%%%%%

\begin{table*}[p]
\centering
\scriptsize
\setlength{\tabcolsep}{4pt}
\renewcommand{\arraystretch}{1.15}

\caption{Integrated Ultrafast Pulse Amplification}
\label{tab:Pulsed_Amp}

\begin{tabular*}{\textwidth}{@{\extracolsep{\fill}} l c c c c c l}
\toprule
\textbf{Active Medium} &
\textbf{WL} &
\textbf{Pulse} &
\textbf{Input} &
\textbf{Rep. rate} &
\textbf{On-chip} &
\textbf{Reference} \\
&
\textbf{(nm)} &
\textbf{energy (nJ)} &
\textbf{width (ps)} &
\textbf{(MHz)} &
\textbf{gain (dB)} &
\\
\midrule

$\mathrm{Yb\!:\!AlO}_{\!x}$ &
1050 &
5.3 &
5.2 &
83.3 &
18.9 &
This work \\
\hline

Ti:Sapphire &
800 &
2.3 &
2.2 &
1.25 &
17.0 &
\citeSI{Vuckovic:2024:Nature} \\

Tm:Al$_{2}$O$_{3}$ &
1815 &
0.095 &
2.5 &
1,000 &
17 &
\citeSI{Herr:2024:Nat_com} \\

Er:Si\textsubscript{3}N\textsubscript{4} &
1558 &
0.131 &
- &
100 &
24 &
\citeSI{kippenberg:2022:Science} \\

Er:Ta$_{2}$O$_{5}$ &
1550 &
7.1$\times\!10^{-4}$ &
38 &
6,000 &
3.8 &
\citeSI{Choudarhary:2025:Elsevier_Optical_Materials} \\

Cr:ZnS CPA\textsuperscript{a} &
2336 &
33.6 &
500 &
70 &
18.8 &
\citeSI{sorokina:2025:OE:CrZnS} \\

III-V TSA\textsuperscript{b} &
780 &
1.78 &
150 &
100 &
20 &
\citeSI{Ye:2006:Optics_letters} \\

III-V QD TSA &
1260 &
$\sim$0.062 &
1.37 &
16,000 &
$\sim$22 &
\citeSI{Weber:2015:Optics_letters} \\

III-V SOA\textsuperscript{c} &
835 &
0.043 &
- &
323 &
- &
\citeSI{delfyett:2005:IEEE:185fs} \\

III-V cascaded SOA XCPA\textsuperscript{d} &
975 &
238 &
6,480 &
57 &
41 (cascaded) &
\citeSI{delfyett:2005:OE:xcpa},\citeSI{delfyett:2005:IEEE:238nJ} \\

\bottomrule
\end{tabular*}

\vspace{0.5em}
\begin{minipage}{\textwidth}
\footnotesize
\textsuperscript{a} CPA: chirped pulse amplification. Depressed cladding buried waveguide in Cr-doped ZnS crystal on Si, inscribed by pulse laser direct writing.
\textsuperscript{b} TSA: tapered semiconductor amplifier.
\textsuperscript{c} SOA: semiconductor optical amplifier.
\textsuperscript{d} XCPA: extreme chirped pulse amplification. Multiple SOAs were cascaded.
\end{minipage}

\end{table*}

\clearpage
\bibliographystyleSI{naturemag}
\bibliographySI{Reference_SI}

\end{document}